\documentclass[10pt,twocolumn,superscriptaddress,english,pra,showpacs,floatfix,aps]{revtex4-2}
\usepackage[utf8]{inputenc}
\usepackage{amsmath}
\usepackage{amssymb}
\usepackage{graphicx}
\usepackage{xspace}
\usepackage{mathtools}
\usepackage{comment}
\usepackage{physics}
\usepackage{bm}
\usepackage{babel}

\makeatletter
 
\usepackage[normalem]{ulem}
\usepackage{xcolor}
\usepackage{soul}
\usepackage[backref=none,
bookmarksnumbered=true,
bookmarks=true,
bookmarksopen=true,
colorlinks=true,
citecolor=blue,
linkcolor=blue,
anchorcolor=green,
urlcolor=blue,unicode=false]{hyperref}

\makeatother

\begin{document}

\preprint{APS/123-QED}

\title{Quantized dynamical pumping via dissipation in a mechanical Thouless pump}

\author{Marius J\"urgensen}
    \email{marius.juergensen@gmail.com}
    \affiliation{Department of Physics, The Pennsylvania State University, University Park, Pennsylvania 16802, USA}%
\author{Mikael C. Rechtsman}%
    \affiliation{Department of Physics, The Pennsylvania State University, University Park, Pennsylvania 16802, USA}%

\date{\today}

\begin{abstract}

Thouless pumps are time-periodic one-dimensional systems that capture the physics of the two-dimensional quantum Hall effect via the quantized pumping of particles under adiabatic modulation. Recent work in photonics has shown that nonlinearity can act to quantize the displacement of light in the form of soliton motion. Here we use a mechanical system -- namely coupled pendulums described by the Frenkel-Kontorova model -- to propose and observe quantized non-adiabatic Thouless pumping using topological kink solitons. The pumping proceeds by a qualitatively different mechanism compared to Thouless' original proposal as the pump is non-adiabatic and dissipation is necessary. In the presence of an additional potential gradient along the pump, we predict and observe the emergence of quantized transport against the pumping direction as a function of the period and show the emergence of a rich plateau structure; this quantization is unique to dissipative systems and cannot be described by the Chern number. Finally, we experimentally demonstrate the robustness of the process by pumping the soliton through a tunable nonlinear defect.

\end{abstract}

\maketitle

\section{Introduction}

In his seminal work, David Thouless proposed that, per period, the pumped electric charge from a spatially and time periodic potential is quantized by the Chern number -- given a Fermi level in the bandgap and adiabatic modulation \cite{thoulessQuantizationParticleTransport1983}. These Thouless pumps have garnered great interest due to their intimate connection with topological phases, and Chern insulators in particular. Specifically, Thouless pumps are dimensionally reduced versions of Chern insulators, for which one wavevector dimension is replaced by a time-periodic modulation. As topological phases are based on generic wave physics, observations are possible in a variety of platforms \cite{citroThoulessPumpingTopology2023} that are governed by different wave equations. For example, Thouless pumps have been observed in photonics \cite{krausTopologicalStatesAdiabatic2012,verbinObservationTopologicalPhase2013}, ultracold atomic systems \cite{lohseThoulessQuantumPump2016, nakajimaTopologicalThoulessPumping2016}, mechanical systems \cite{grinbergRobustTemporalPumping2020}, spins \cite{schweizerSpinPumpingMeasurement2016}, and their interplay with disorder \cite{cerjanThoulessPumpingDisordered2020,nakajimaCompetitionInterplayTopology2021}, nonlinearity \cite{jurgensenQuantizedNonlinearThouless2021,jurgensenQuantizedFractionalThouless2023}, inter-particle interactions \cite{walterQuantizationItsBreakdown2023,viebahnInteractioninducedChargePumping2023}, synthetic dimensions \cite{martinTopologicalFrequencyConversion2017}, and dissipation \cite{fedorovaObservationTopologicalTransport2020,dreonSelfoscillatingPumpTopological2022,sridharQuantizedTopologicalPumping2024} have been examined.

Topological protection has been studied in a wide range of different contexts, including solid-state physics \cite{hasanColloquiumTopologicalInsulators2010}, cold atoms in lattices \cite{cooperTopologicalBandsUltracold2019}, photonics \cite{ozawaTopologicalPhotonics2019}, and mechanical systems \cite{nashTopologicalMechanicsGyroscopic2015, huberTopologicalMechanics2016}. Mechanical systems offer unique advantages and disadvantages: on one hand, dissipation is often much greater as compared to optical or cold atomic systems, but on the other hand, a high degree of nonlinearity is straightforward to achieve. In photonics or ultracold atomic gases, nonlinearities are used to describe interacting bosons in the mean-field limit. In other words, nonlinearities capture the consequences of inter-particle interactions in systems with many bosons per site, completely agnostic of the underlying experimental system, that can be mechanical, acoustic, optical or otherwise.




Here, we propose and observe quantized Thouless pumping of topological kink solitons -- nonlinear eigenstates that cannot be destroyed due to their topological charge -- using a mechanical system of coupled pendulums whose nonlinear dynamics are described by a discrete sine-Gordon equation (also known as the Frenkel-Kontorova model). In order to realize the pump dynamics, we temporally modulate the effective gravitational acceleration for the pendulums using fans mounted beneath them. When adding a potential gradient to the model (which is analogous to an electric field imposed upon electrons), the model undergoes a transition determined by the cycle period and we observe the emergence of a friction-induced quantization of transport opposite the direction of the pumping sequence, without the need for adiabaticity. Hence, this quantization has no analogue in traditional wave-based platforms such as optical and electronic systems. Finally, we test the robustness of nonlinear Thouless pumping against a linear and nonlinear defect and show that while small defects do not break quantization, stronger defects can.

\begin{figure*}[htbp]
    \includegraphics[width=\textwidth]{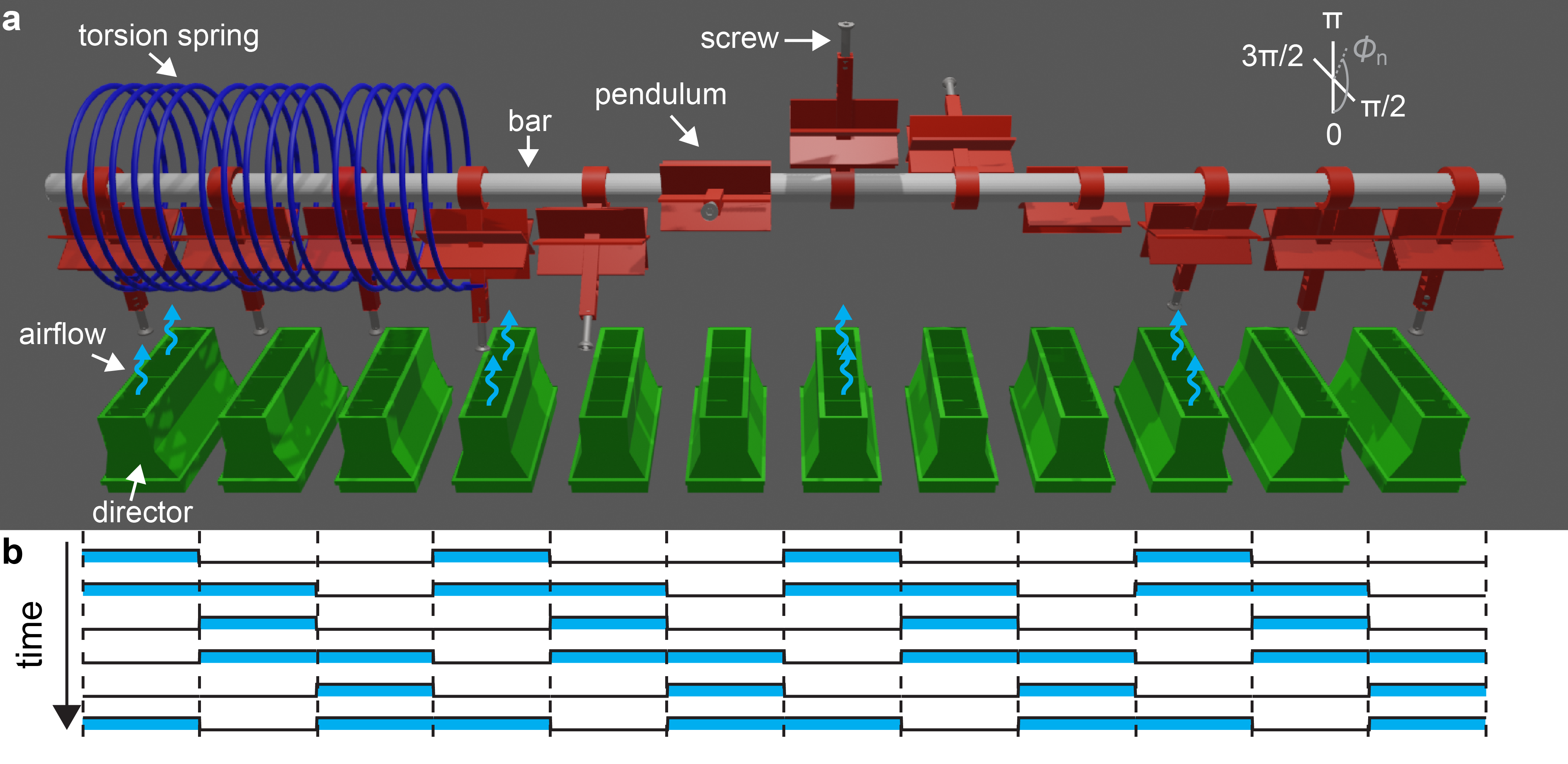}
    \caption{\textbf{Mechanical nonlinear Thouless pumping model.} \textbf{a} Identical pendulums (red) are mounted on a rod (grey) and connected via torsion springs (dark blue). For better visibility, only three springs are shown. Screws (grey) serve as additional weights for the pendulums. Fans are mounted below the pendulums, whose airflow (light blue) is directed vertically upwards using directors (green). The pendulums show a "windmill" structure to increase the effect of the air flow. The rotation of the pendulums shows a kink soliton and the coordinate system is defined in the top right. \textbf{b} The Thouless pumping sequence. The fans as shown in \textbf{a} are subsequently switched on and off. A forward pumping sequence is shown, which spatially repeats every three sites and is analogous to a three site Aubry-André-Harper model.}
    \label{Fig:model}
\end{figure*}

\begin{figure*}[htbp]
    \includegraphics[width=\textwidth]{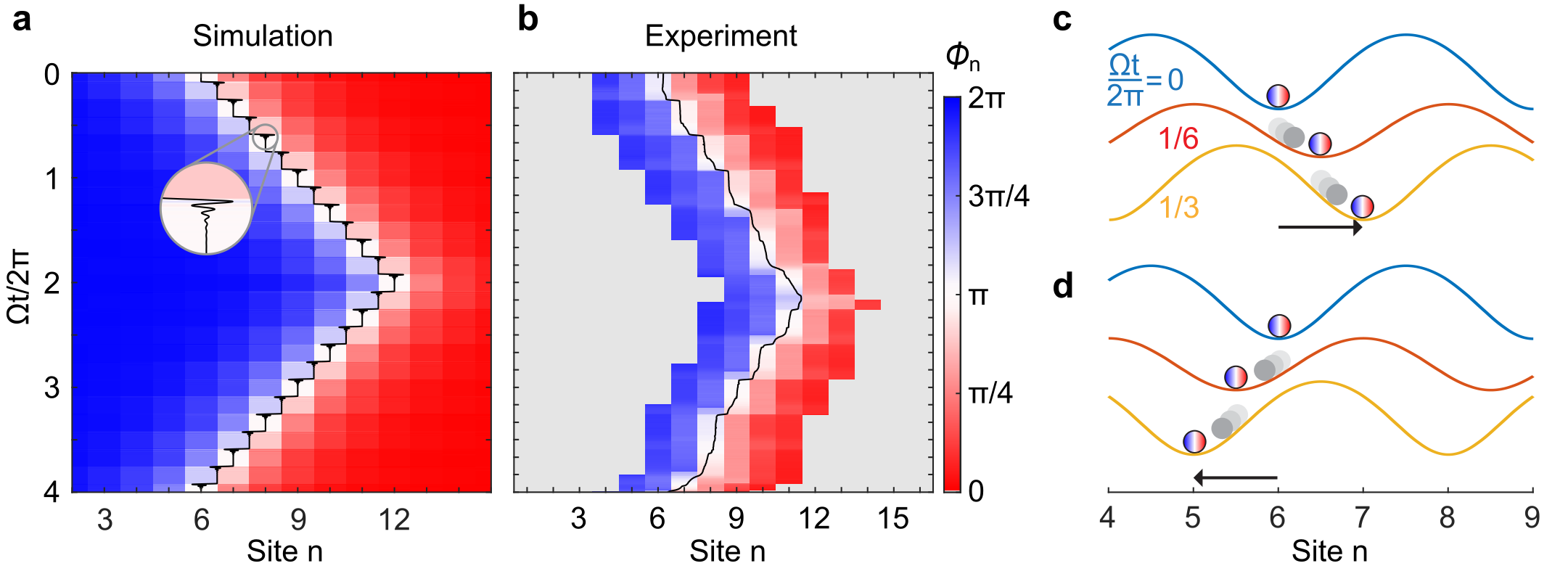}
    \caption{\textbf{Observation of quantized forward and backward kink soliton pumping.} \textbf{a} Simulated propagation of a kink soliton for two forward periods and two backwards periods showing transport by plus or minus one unit cell (three sites) per period. The black line shows the center of mass position of the kink soliton. \textbf{b} Experimental observation of kink soliton pumping. Grey values denote pendulums that are not measurable by a camera with a top view. \textbf{c} Theoretical effective energy landscape calculated for the respective instantaneous kink soliton. Blue, red, yellow are calculated for different points in time during one pumping cycle ($\Omega t/2\pi$=0, 1/6, 1/3). The soliton, symbolized by the ball, follows the potential minimum. Importantly, this is only possible with friction. \textbf{d} Similar to \textbf{c} but for reverse pumping. Here, we use $T$=60\,s, $\alpha$=2.4/s, $k$=2.5\,mNm/rad and no additional weight on the pendulums.}
    \label{Fig:FB}
\end{figure*}

\section{Description of the model}
Our mechanical system is shown in Fig. 1a and consists of nominally identical, equidistantly spaced pendulums that can rotate around a horizontal bar using bearings (not shown) to reduce friction. Neighboring pendulums are coupled using torsion springs forming a prototypical discrete sine-Gordon model (also known as Frenkel-Kontorova model \cite{braunNonlinearDynamicsFrenkel1998}, originally developed to describe dislocations in crystal lattices). At the end of each pendulum, a screw and additional nuts can be attached to change the pendulums’ masses. Later, this will be utilized to create a defect in the pendulum lattice. Furthermore, beneath each pendulum, a fan can create a vertical airflow that counteracts the gravitational force on the pendulum, changing its potential energy.  It is through these fans that the Thouless pumping sequence is implemented. To reach the necessary modulation strength using the fans, all pendulums are spatially extended using a ``windmill'' structure that makes the force from the fan independent of the pendulum's rotation angle. Pendulums, springs and the fan airflow directors are 3D printed using polylactic acid (PLA) (more information can be found in Supplementary Section A).

The dynamics of our system are given by 
\begin{align}
    \ddot{\phi}_n &= -D_n(t) \sin\phi_n + K \left( \phi_{n+1}+\phi_{n-1}-2 \phi_n \right) - \alpha \dot{\phi}_n, 
\end{align}
where $\phi_n$ is the rotation angle of pendulum $n$, the dot symbolizes a derivative with respect to time $t$, $D_n(t)$ is related to the potential energy of the pendulum that can change over time when the airflow is changed, $K=k/I$ is related to the strength of the energy stored in the torsion springs with spring constant $k$ and moment of inertia $I$, and $\alpha$ describes the strength of friction. A derivation of this equation from the respective Lagrangian is given in Supplementary Information Section D.

The discrete sine-Gordon equation (Eq. 1) allows for the formation of solitons that are localized nonlinear eigenstates. In this work, we are only concerned with kink solitons that connect two different degenerate ground states, for example $\phi_n=0$ and $\phi_n = 2\pi$ for all $n$. The solitons are topologically protected in the sense that they wind (in this case) through $2\pi$ radians, meaning that they cannot be smoothly deformed back to the constant-amplitude state. An example of such a kink soliton is shown in Fig. 1a. Assuming a sample with closed boundary conditions such that there is a single turn throughout the whole system, single kink solitons can neither be destroyed nor created. In the continuum description, kink solitons have exact analytical solutions, can be Lorentz boosted, and have a velocity-dependent width \cite{braunNonlinearDynamicsFrenkel1998}. In discrete systems, kink soliton must overcome the Peierls-Nabarro barrier \cite{kivsharPeierlsNabarroPotentialBarrier1993} in order to move in the lattice, which leads to soliton friction and finally brings the soliton to a stop in static discrete lattices \cite{braunNonlinearDynamicsFrenkel1998}.

In this work, solitons are not boosted in order to move by giving them some initial velocity, but rather are transported by time-periodically changing (with period $T$) the Hamiltonian of the system creating a Thouless pump. To this end, we modulate the airflow as a function of time by switching the fans on and off. As can be seen in Fig. 1b, one period is divided into six equally-long time intervals in which either one or two fans per unit cell are constantly switched on, while the others are switched off. Due to the "windmill" structure of the pendulums, the airflow creates an angle-independent upwards force that counteracts the gravitational force. Consequently, an effectively lower gravitation acceleration acts on those pendulums (for more information, see Supplementary Section E). In our Thouless pumping model, every third fan is driven identically. Hence our model has a three-site unit cell and our driving sequence is analogous to the on-diagonal version of the Aubry-André-Harper model, as known from condensed matter physics \cite{harperSingleBandMotion1955, aubryAnalyticityBreakingAnderson1980} where it has been shown to lead to quantized transport of electrons in the adiabatic limit ($T \rightarrow \infty$).

In strong contrast to conventional Thouless pumps, for which quantization occurs given a slowly changing Hamiltonian, and becomes exact in the adiabatic limit, here the fans are switched on and off instantaneously. Thus, our modulation sequence has no adiabatic limit, which shows that the original topological theory of Thouless pumping is insufficient to describe our system. Furthermore, no topological invariant is known for our system, although we clearly observe regimes with quantized behavior. As we will show below, dissipation (in the form of friction) is crucial to enable this quantization. Our results are model-independent and also apply to Thouless pumps with time-continuous modulations (for a sinusoidally-modulated Rice-Mele model see Supplementary Information Section G) in the presence of dissipation. 

Importantly, for our system, dissipation has a different effect than in the Schr\"odinger case, in the sense that it does not lead to a diminishment of the soliton or its amplitude, since the soliton is topologically protected. It only effects the velocity of the soliton. In the following, we will describe three separate experiments showing quantized pumping of topological kink solitons.



First, using the pumping sequence described above, we show quantized kink soliton pumping. To this end, we initially prepare a kink soliton by turning a pendulum at the boundary by 2$\pi$; subsequently the kink soliton can be moved into the bulk of the system, where it is stable. We then run the pumping sequence for two periods (pumping to the right), before reversing the sequence for two additional periods (pumping to the left). Figure 2a shows the calculated behavior. Blue/red denote the pendulums pointing downwards. The kink soliton is represented by the winding of the pendulums through a full $2\pi$ rotation angle. After each period, the soliton (with center-of-mass shown in black) is transported by three sites (one unit cell) to the right (for the forward sequence) or to the left (by the reverse sequence). The corresponding experiment is shown in Fig. 2b. We evaluate the position of the soliton using a camera above the pendulums that faces downwards, as shown in Supplementary Video 1. In this mode, reliable tracking only works for angles $25\lessapprox \phi_n \lessapprox 335$ (see also Supplementary Information Section C), for which the pendulums are visible from the top. The center of the experimental soliton is extracted from a fit using the wavefunction of a kink soliton in the continuum. Due to unavoidable fabrication disorder (e.g. variations from site to site) and friction, the soliton does not match the simulations perfectly. But it clearly shows the same pumping behavior to the right and left, confirming the quantized pumping behavior.

\begin{figure*}[htp]
    \includegraphics[width=\textwidth]{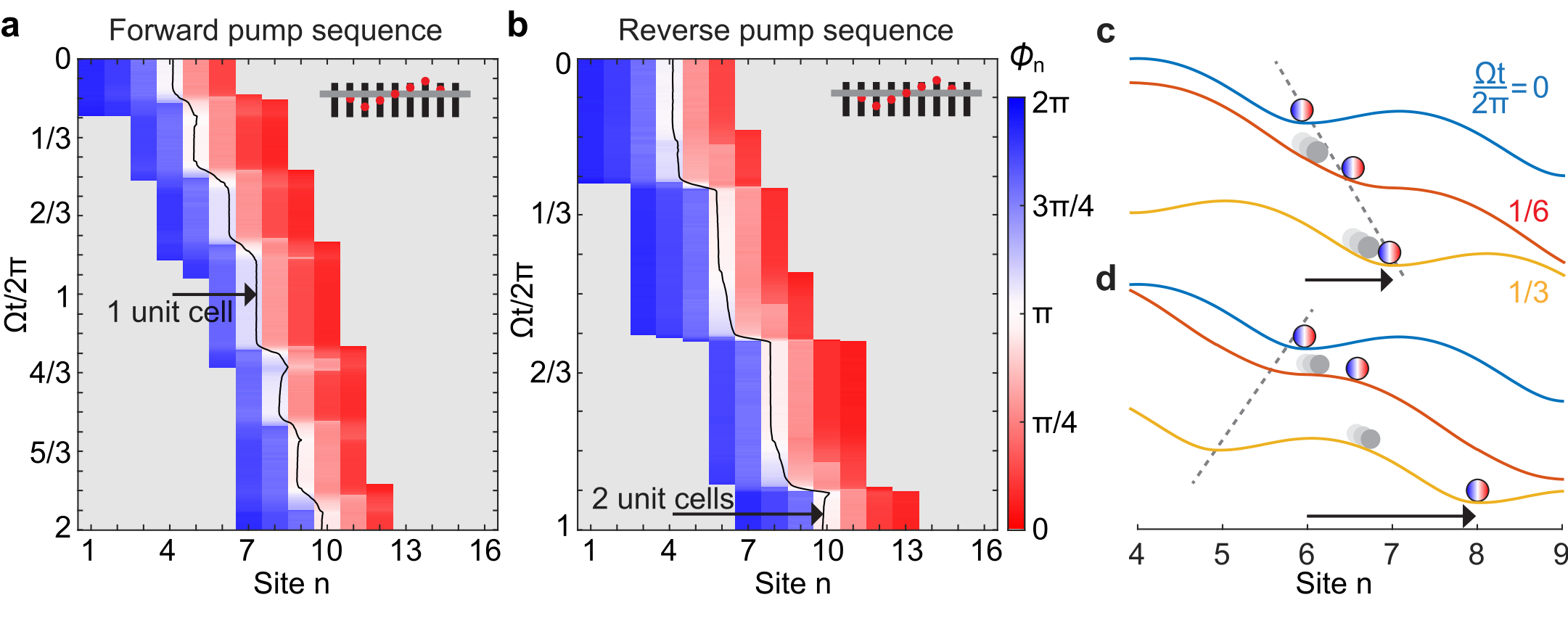}
    \caption{\textbf{Observation of friction induced reverse Thouless pumping in the presence of a potential gradient.} \textbf{a} In forward pumping sequence, the kink soliton is displaced by one unit cell (three sites) per period in the pumping direction. Here, two periods are shown. Black line denotes the center of mass position of the kink soliton. The inset illustrates the shifted position of the fans (black) in relation to the pendulums attached to the bar (grey). \textbf{b} Similar to \textbf{a}, but for the reverse pumping sequence. In the presence of the potential gradient, the kink soliton is moved by two unit cells (six sites) against the pumping direction (but along the potential gradient). \textbf{c,d} Similar to Fig. 2c,d, but including the influence of the potential gradient. Clearly, no potential minimum exists at $\Omega t/2\pi=1/6$, implying the sequence cannot be performed adiabatically. Friction is necessary to ``cool down" the soliton into the local potential minimum. The grey dashed line symbolized the position of the soliton in a system without potential gradient. For \textbf{c},\textbf{d}, we use an imbalance of airflow given by $\beta$=0.35.}
    \label{Fig:withEfield}
\end{figure*}

While quantized soliton pumping has previously been observed by the authors in photonics \cite{jurgensenQuantizedNonlinearThouless2021}, this crucially relied on adiabaticity and low loss. In other words, the relevant dissipation length was larger than the period. Here, we show quantization in a dynamical (non-adiabatic) pump {\it induced} by dissipation. Indeed, in our system (linear) oscillations die out after $1/\alpha<0.5$\,s, which is a small fraction of our Thouless pumping period of $T$=60\,s. This means that the quantization mechanism is fundamentally different. To understand the quantization we use the variational method, which is justified as the kink soliton represents the lowest energy state when the pendulums are constrained to have one full $2\pi$ winding. We use the analytical solution of the kink soliton in the continuum, $\psi(x)=4\arctan ( e^{w (x-x_0)+\delta})$, as an ansatz for the wavefunction, and vary the width $w$ for fixed positions $x_0$. This variational approach drastically decreases the phase space (from originally $2N$ to 2 dimensions, where $N$ is the number of pendulums) and allows not only for the calculation of the ground state, but also an approximate energy landscape (for more information see Supplementary Information Section D). In other words, we calculate the effective potential landscape in which the soliton resides. This is shown in Fig. 2c and d, for a forward and reverse pumping sequence, respectively. 

Suppose the soliton starts in a local potential minimum (blue curve, Fig. 2c,d). At time $\Omega t=1/6$, additional fans are switched on and the Hamiltonian changes leading to a new potential energy landscape (red curve). As the soliton is not in a local potential minimum anymore, the potential exerts a force onto the soliton and accelerates it towards decreasing potential. Although being an intricate nonlinear object, in this simplified picture, the soliton can be regarded as a ball, rolling down a hill. Without friction, the soliton would oscillate around the local minimum, but in the presence of dissipation (friction), the soliton converges into the local minimum (this behavior can be clearly seen in the inset of Fig. 2a). Hence, over the course of the Thouless pumping cycle, we find numerically that the soliton follows the local potential minimum with a displacement that is equal to the Chern number of the lowest band for the corresponding linear model with no dissipation (see Supplementary Information Section F).  


In a second experiment we add an effective potential gradient along the pumping direction. In an electronic system, this would be equivalent to adding a constant electric field. For kink solitons we achieve this by horizontally displacing the fans below the pendulums, orthogonal to the bar (see insets in Fig. 3a and b). As a consequence, pendulums at rotation angles $0 < \phi_n < \pi$, feel a different upwards force compared to pendulums at rotation angles  $\pi < \phi_n < 2\pi$. We quantify the difference using the parameter $\beta$. Importantly, this displacement leads to an effective gradient in the interplay with kink solitons. Furthermore, kink and anti-kink solitons experience oppositely oriented potential gradients. More information and a formal derivation can be found in Supplementary Information Section E.

\begin{figure*}[tbh]
    \includegraphics[]{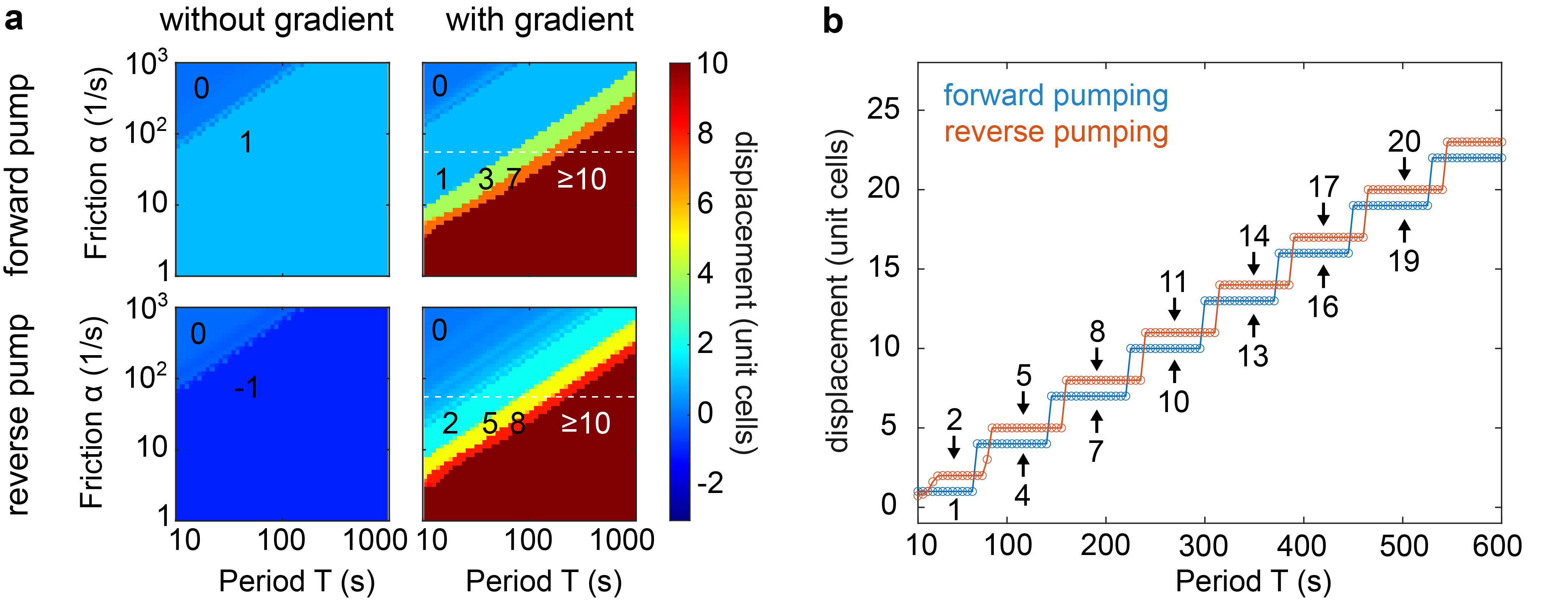}
    \caption{\textbf{Phase diagrams.} \textbf{a} Phase diagrams for a Thouless pump without and with a potential gradient, for forward and reverse pumping direction. The numbers describe the displacement of the soliton per period in units of the unit cell length. Multiple quantized regimes are visible due to an interplay between non-adiabaticity, friction and the potential gradient. \textbf{b} Slice through the phase diagrams in \textbf{a} (denoted by the white dashed line), showing multiple plateaux with a quantized displacement of $1+3m$ and $2+3m$ for forward pumping (blue) and reverse pumping (red), respectively. The phase diagrams have been calculated using $\beta$=0 and $\beta$=0.35, for the case without and with the gradient, respectively, $k=2.7$mNm/rad and no additional nuts attached to the pendulums.}
    \label{Fig:PhaseDiagram}
\end{figure*}

We choose to use a kink soliton and create a potential gradient descending from left to right. Under these fixed conditions, we run our experiment using two different protocols. In the first protocol, depicted in Fig. 3a, we use a forward pumping scheme and pump in the direction of \textit{decreasing} potential. In the second protocol, shown in Fig. 3b, we reverse the pumping sequence, that is, we pump in the direction of \textit{increasing} potential. No other changes are made, i.e. we use the same starting position for the kink soliton. In the first variant (forward pump sequence), we observe that the pumping quantization is identical to the case without potential gradient (as shown in Fig. 2). This is visible by looking at the center-of-mass trajectory of the soliton (which is shown for two periods in Fig. 3a). After each period, the soliton is transported by one unit cell (three sites) to the right, corresponding to a Chern number of 1. Small deviations stem from disorder which shifts the position of the soliton locally. In the second variant (reverse pump sequence) we observe a quantized pumping by two unit cells (six sites) per period, also to the right. This is in stark contrast to the case without potential gradient, where the soliton is transported by one unit cell to the left.

This behavior can be understood from Fig. 3c,d, where we show the effective potential landscape for the kink soliton, given the potential gradient as a result of the displaced pendulums. Essentially, Figs. 3c and d, are analogous to Figs. 2c and d with the addition of a linear potential gradient. As can be seen from the potential landscape at $\Omega t=1/6$ (red line), there are intervals in the pumping sequence with no local potential minima. Hence, in a frictionless adiabatic system, the soliton would accelerate indefinitely. But if non-adiabaticity and friction act in such a way that the soliton is captured by the next local minimum in a subsequent time interval (e.g. at $\Omega t = 1/3$; yellow line), quantized pumping can be restored. Notably, during the reverse pumping sequence, the capturing potential minimum is in the direction of the the descending gradient, one unit cell to the right of the potential minimum without the gradient. Hence, instead of pumping by one unit cell to the left, it pumps by two unit cells to the right (after one period). As we will show in the following, the number of pumped cells is dependent on the period. Hence, the observed quantization cannot be explained using Thouless' original argument, for which adiabaticity is crucial. Instead, these observations suggest a novel quantization mechanism that relies on the interplay of friction, non-adiabaticity and a potential gradient.


Since our analysis suggests that the quantized pumping dynamics are determined by the interplay between friction and the degree of adiabaticity, we calculate phase diagrams for forward and reverse pumping, with and without a potential gradient. These results are shown in Fig. 4a and b. As the dynamical solutions are not known {\it a-priori}, our simulations start using an instantaneous soliton guess, and we run the simulations for multiple periods. We then evaluate the displacement during the third period. Without a potential gradient, we observe two regimes of different quantization: at low friction or long periods, the soliton is pumped by an integer in the direction of the pumping sequence, while when friction dominates, the transport is zero. Apart from the transport orientation, forward and backwards pumping are identical.

The addition of a potential gradient modifies the picture. In both cases, for the forward and reverse pumping sequence, there is only pumping in the rightward direction, the direction of decreasing potential. With decreasing friction or increasing period a rich plateaux structure emerges, described by integers that are associated with the number and direction of unit cells that the soliton moves per pump cycle. In Fig. 4b, we show the occurrence of these plateaux, with displacement values of $1+3m$ and $2+3m$ ($m \in \mathbb{N}$) for forward (blue) and backwards (red) pumping, respectively. Importantly, these new regimes necessarily require dissipation and non-adibaticity and are not equal to the Chern numbers of the gradient-free system. For a more intuitive understanding, we show the movement of forward-pumped solitons over one period ($T=60\,$s) in their respective effective potential landscape in Supplementary Animation 1. Choosing $\alpha=\{16,36,100\}$/s, we show quantized displacements of $\{1,4,7\}$ unit cells, respectively. We note that, as the onset of the quantized re is strongly sensitive to the chosen parameters (e.g., $\beta$, $k$), we do not expect perfect quantitative agreement with our experimental observation. Nevertheless, qualitatively, we see in Fig. 4b that there is a regime for which forward pumping results in +1 unit cell and reverse pumping in +2 unit cells of displacement, albeit at a higher friction. Importantly, the quantized regimes are robust to small changes of the friction coefficient, the period of the pump and the strength of the potential gradient (see also Supplementary Information Section H).



\begin{figure}[htbp]
    \includegraphics[]{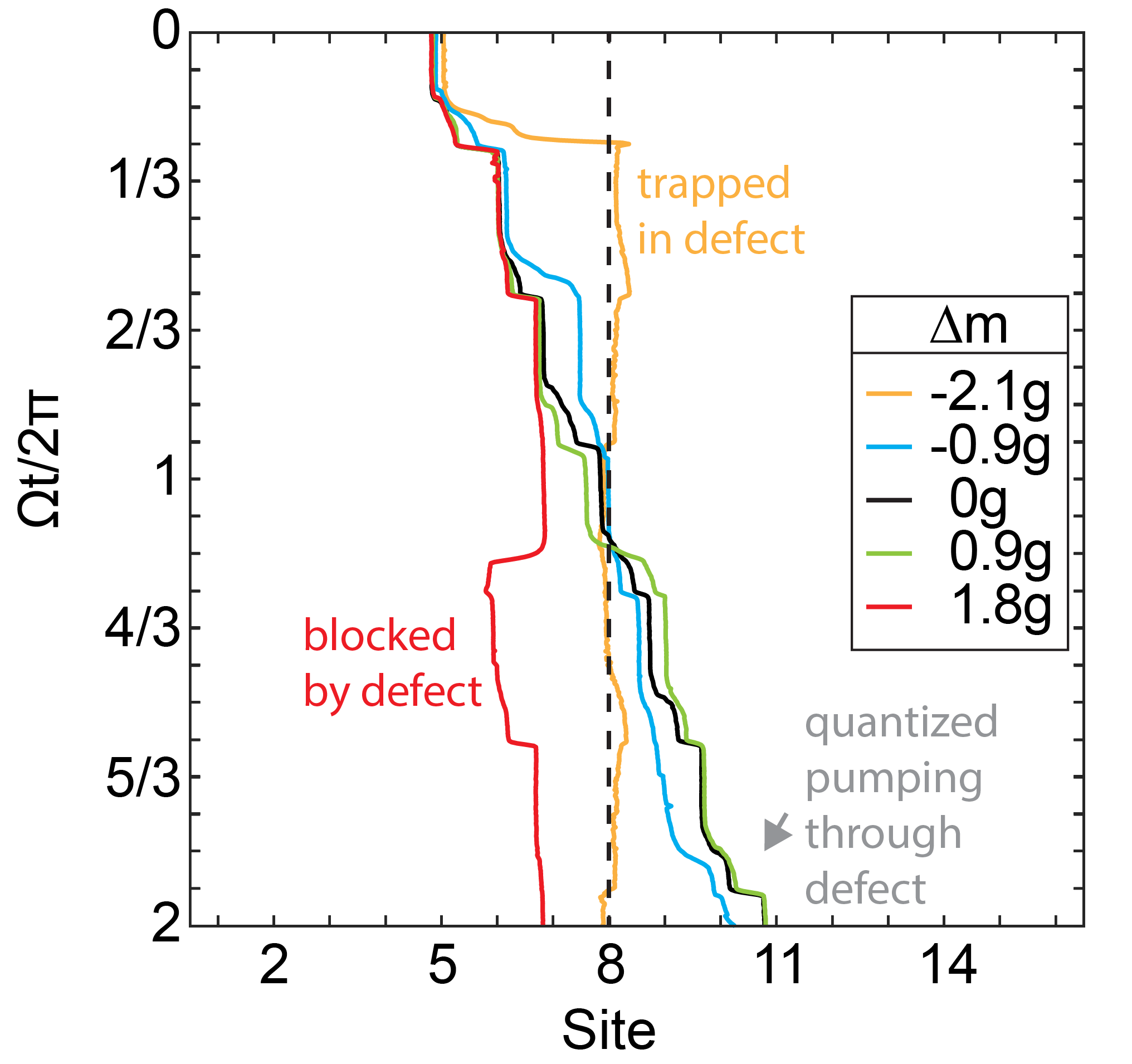}
    \caption{\textbf{Observation of protection against attractive and repulsive nonlinear defects.} Experimentally observed center of mass position of kink solitons over two periods. The defect is created by changing the mass ($\Delta m$) of the pendulum of site $n=8$ and symbolized by the dashed line. Without a defect (black), and for small defect mass increase/decrease (green/blue) pumping is quantized. For larger defect mass increase/decrease (red/yellow) the defect becomes strongly attractive/repulsive and quantization breaks down. In this experiment all pendulums (except the defect) have one nut attached. The yellow curve symbolizes a pendulum without a nut and without a screw. From blue to red one nut each is added.}
    \label{Fig:Defect}
\end{figure}

Third, and finally, we experimentally test the robustness of the quantization for our pumping scheme for kink solitons. To this end, we create a single defect in the center of the system (on site number 8) by decreasing/increasing the mass of this pendulum. Importantly, due to the sine-term in Eq. (1), changing the effective gravitational acceleration of a pendulum forms an effective linear as well as a nonlinear defect. We start with a kink soliton centered three sites to the left of the defect, far enough away such that the influence of the defect on the kink soliton position is negligible. We then run two forward pumping cycles, seeking to pump the soliton by six sites to the right and across the defect.

We observe that for small defects, the soliton can be pumped through the defect. It restores its quantized behavior once it is sufficiently distant from the defect. With increasing defect mass, we observe that the trajectory of the soliton "lags" behind the defect-free case, while crossing the defect site, until it is blocked by the defect at even lower defect mass. For increasing defect mass we observe that the soliton tends be attracted to the defect when in its vicinity. For sufficiently strong defects it cannot be pumped through anymore and instead is trapped on the defect site. This shows that nonlinear Thouless pumping of solitons is protected against small disorder, which may itself even be nonlinear in nature.

\section{Summary \& Outlook}
In summary, we have observed quantized pumping of topological solitons in a dissipative system of coupled pendulums with modulated airflow, as described mathematically by the Frenkel-Kontorova model. Our system highlights the classical wave character of Thouless pumping of solitons. In contrast to electronic Thouless pumps, quantization occurs due to the interplay of non-adiabaticity and dissipation. The addition of a potential gradient, acting as a force on the soliton, allowed us to observe the emergence of a novel quantized regime. Theoretically, we find the emergence of a rich plateaux structure. Finally, we showed that nonlinear Thouless pumping is protected against small linear and nonlinear defects, before quantization breaks down for stronger defects.


\begin{acknowledgments}
We acknowledge the support of the AFOSR MURI program under agreement number FA9550-22-1-0339, the ONR MURI program, under agreement number N00014-20-1-2325 as well as the ONR under agreement number N00014-23-1-2102. We also acknowledge Madeline Halkowski for providing us with the Atmega module.
\end{acknowledgments}

\bibliography{bibliography}

\begin{thebibliography}{29}%
\makeatletter
\providecommand \@ifxundefined [1]{%
 \@ifx{#1\undefined}
}%
\providecommand \@ifnum [1]{%
 \ifnum #1\expandafter \@firstoftwo
 \else \expandafter \@secondoftwo
 \fi
}%
\providecommand \@ifx [1]{%
 \ifx #1\expandafter \@firstoftwo
 \else \expandafter \@secondoftwo
 \fi
}%
\providecommand \natexlab [1]{#1}%
\providecommand \enquote  [1]{``#1''}%
\providecommand \bibnamefont  [1]{#1}%
\providecommand \bibfnamefont [1]{#1}%
\providecommand \citenamefont [1]{#1}%
\providecommand \href@noop [0]{\@secondoftwo}%
\providecommand \href [0]{\begingroup \@sanitize@url \@href}%
\providecommand \@href[1]{\@@startlink{#1}\@@href}%
\providecommand \@@href[1]{\endgroup#1\@@endlink}%
\providecommand \@sanitize@url [0]{\catcode `\\12\catcode `\$12\catcode `\&12\catcode `\#12\catcode `\^12\catcode `\_12\catcode `\%12\relax}%
\providecommand \@@startlink[1]{}%
\providecommand \@@endlink[0]{}%
\providecommand \url  [0]{\begingroup\@sanitize@url \@url }%
\providecommand \@url [1]{\endgroup\@href {#1}{\urlprefix }}%
\providecommand \urlprefix  [0]{URL }%
\providecommand \Eprint [0]{\href }%
\providecommand \doibase [0]{https://doi.org/}%
\providecommand \selectlanguage [0]{\@gobble}%
\providecommand \bibinfo  [0]{\@secondoftwo}%
\providecommand \bibfield  [0]{\@secondoftwo}%
\providecommand \translation [1]{[#1]}%
\providecommand \BibitemOpen [0]{}%
\providecommand \bibitemStop [0]{}%
\providecommand \bibitemNoStop [0]{.\EOS\space}%
\providecommand \EOS [0]{\spacefactor3000\relax}%
\providecommand \BibitemShut  [1]{\csname bibitem#1\endcsname}%
\let\auto@bib@innerbib\@empty
\bibitem [{\citenamefont {Thouless}(1983)}]{thoulessQuantizationParticleTransport1983}%
  \BibitemOpen
  \bibfield  {author} {\bibinfo {author} {\bibfnamefont {D.~J.}\ \bibnamefont {Thouless}},\ }\bibfield  {title} {\bibinfo {title} {Quantization of particle transport},\ }\href {https://doi.org/10.1103/PhysRevB.27.6083} {\bibfield  {journal} {\bibinfo  {journal} {Physical Review B}\ }\textbf {\bibinfo {volume} {27}},\ \bibinfo {pages} {6083} (\bibinfo {year} {1983})}\BibitemShut {NoStop}%
\bibitem [{\citenamefont {Citro}\ and\ \citenamefont {Aidelsburger}(2023)}]{citroThoulessPumpingTopology2023}%
  \BibitemOpen
  \bibfield  {author} {\bibinfo {author} {\bibfnamefont {R.}~\bibnamefont {Citro}}\ and\ \bibinfo {author} {\bibfnamefont {M.}~\bibnamefont {Aidelsburger}},\ }\bibfield  {title} {\bibinfo {title} {Thouless pumping and topology},\ }\href {https://doi.org/10.1038/s42254-022-00545-0} {\bibfield  {journal} {\bibinfo  {journal} {Nature Reviews Physics}\ }\textbf {\bibinfo {volume} {5}},\ \bibinfo {pages} {87} (\bibinfo {year} {2023})}\BibitemShut {NoStop}%
\bibitem [{\citenamefont {Kraus}\ \emph {et~al.}(2012)\citenamefont {Kraus}, \citenamefont {Lahini}, \citenamefont {Ringel}, \citenamefont {Verbin},\ and\ \citenamefont {Zilberberg}}]{krausTopologicalStatesAdiabatic2012}%
  \BibitemOpen
  \bibfield  {author} {\bibinfo {author} {\bibfnamefont {Y.~E.}\ \bibnamefont {Kraus}}, \bibinfo {author} {\bibfnamefont {Y.}~\bibnamefont {Lahini}}, \bibinfo {author} {\bibfnamefont {Z.}~\bibnamefont {Ringel}}, \bibinfo {author} {\bibfnamefont {M.}~\bibnamefont {Verbin}},\ and\ \bibinfo {author} {\bibfnamefont {O.}~\bibnamefont {Zilberberg}},\ }\bibfield  {title} {\bibinfo {title} {Topological {{States}} and {{Adiabatic Pumping}} in {{Quasicrystals}}},\ }\href {https://doi.org/10.1103/PhysRevLett.109.106402} {\bibfield  {journal} {\bibinfo  {journal} {Physical Review Letters}\ }\textbf {\bibinfo {volume} {109}},\ \bibinfo {pages} {106402} (\bibinfo {year} {2012})}\BibitemShut {NoStop}%
\bibitem [{\citenamefont {Verbin}\ \emph {et~al.}(2013)\citenamefont {Verbin}, \citenamefont {Zilberberg}, \citenamefont {Kraus}, \citenamefont {Lahini},\ and\ \citenamefont {Silberberg}}]{verbinObservationTopologicalPhase2013}%
  \BibitemOpen
  \bibfield  {author} {\bibinfo {author} {\bibfnamefont {M.}~\bibnamefont {Verbin}}, \bibinfo {author} {\bibfnamefont {O.}~\bibnamefont {Zilberberg}}, \bibinfo {author} {\bibfnamefont {Y.~E.}\ \bibnamefont {Kraus}}, \bibinfo {author} {\bibfnamefont {Y.}~\bibnamefont {Lahini}},\ and\ \bibinfo {author} {\bibfnamefont {Y.}~\bibnamefont {Silberberg}},\ }\bibfield  {title} {\bibinfo {title} {Observation of {{Topological Phase Transitions}} in {{Photonic Quasicrystals}}},\ }\href {https://doi.org/10.1103/PhysRevLett.110.076403} {\bibfield  {journal} {\bibinfo  {journal} {Physical Review Letters}\ }\textbf {\bibinfo {volume} {110}},\ \bibinfo {pages} {076403} (\bibinfo {year} {2013})}\BibitemShut {NoStop}%
\bibitem [{\citenamefont {Lohse}\ \emph {et~al.}(2016)\citenamefont {Lohse}, \citenamefont {Schweizer}, \citenamefont {Zilberberg}, \citenamefont {Aidelsburger},\ and\ \citenamefont {Bloch}}]{lohseThoulessQuantumPump2016}%
  \BibitemOpen
  \bibfield  {author} {\bibinfo {author} {\bibfnamefont {M.}~\bibnamefont {Lohse}}, \bibinfo {author} {\bibfnamefont {C.}~\bibnamefont {Schweizer}}, \bibinfo {author} {\bibfnamefont {O.}~\bibnamefont {Zilberberg}}, \bibinfo {author} {\bibfnamefont {M.}~\bibnamefont {Aidelsburger}},\ and\ \bibinfo {author} {\bibfnamefont {I.}~\bibnamefont {Bloch}},\ }\bibfield  {title} {\bibinfo {title} {A {{Thouless}} quantum pump with ultracold bosonic atoms in an optical superlattice},\ }\href {https://doi.org/10.1038/nphys3584} {\bibfield  {journal} {\bibinfo  {journal} {Nature Physics}\ }\textbf {\bibinfo {volume} {12}},\ \bibinfo {pages} {350} (\bibinfo {year} {2016})}\BibitemShut {NoStop}%
\bibitem [{\citenamefont {Nakajima}\ \emph {et~al.}(2016)\citenamefont {Nakajima}, \citenamefont {Tomita}, \citenamefont {Taie}, \citenamefont {Ichinose}, \citenamefont {Ozawa}, \citenamefont {Wang}, \citenamefont {Troyer},\ and\ \citenamefont {Takahashi}}]{nakajimaTopologicalThoulessPumping2016}%
  \BibitemOpen
  \bibfield  {author} {\bibinfo {author} {\bibfnamefont {S.}~\bibnamefont {Nakajima}}, \bibinfo {author} {\bibfnamefont {T.}~\bibnamefont {Tomita}}, \bibinfo {author} {\bibfnamefont {S.}~\bibnamefont {Taie}}, \bibinfo {author} {\bibfnamefont {T.}~\bibnamefont {Ichinose}}, \bibinfo {author} {\bibfnamefont {H.}~\bibnamefont {Ozawa}}, \bibinfo {author} {\bibfnamefont {L.}~\bibnamefont {Wang}}, \bibinfo {author} {\bibfnamefont {M.}~\bibnamefont {Troyer}},\ and\ \bibinfo {author} {\bibfnamefont {Y.}~\bibnamefont {Takahashi}},\ }\bibfield  {title} {\bibinfo {title} {Topological {{Thouless}} pumping of ultracold fermions},\ }\href {https://doi.org/10.1038/nphys3622} {\bibfield  {journal} {\bibinfo  {journal} {Nature Physics}\ }\textbf {\bibinfo {volume} {12}},\ \bibinfo {pages} {296} (\bibinfo {year} {2016})}\BibitemShut {NoStop}%
\bibitem [{\citenamefont {Grinberg}\ \emph {et~al.}(2020)\citenamefont {Grinberg}, \citenamefont {Lin}, \citenamefont {Harris}, \citenamefont {Benalcazar}, \citenamefont {Peterson}, \citenamefont {Hughes},\ and\ \citenamefont {Bahl}}]{grinbergRobustTemporalPumping2020}%
  \BibitemOpen
  \bibfield  {author} {\bibinfo {author} {\bibfnamefont {I.~H.}\ \bibnamefont {Grinberg}}, \bibinfo {author} {\bibfnamefont {M.}~\bibnamefont {Lin}}, \bibinfo {author} {\bibfnamefont {C.}~\bibnamefont {Harris}}, \bibinfo {author} {\bibfnamefont {W.~A.}\ \bibnamefont {Benalcazar}}, \bibinfo {author} {\bibfnamefont {C.~W.}\ \bibnamefont {Peterson}}, \bibinfo {author} {\bibfnamefont {T.~L.}\ \bibnamefont {Hughes}},\ and\ \bibinfo {author} {\bibfnamefont {G.}~\bibnamefont {Bahl}},\ }\bibfield  {title} {\bibinfo {title} {Robust temporal pumping in a magneto-mechanical topological insulator},\ }\href {https://doi.org/10.1038/s41467-020-14804-0} {\bibfield  {journal} {\bibinfo  {journal} {Nature Communications}\ }\textbf {\bibinfo {volume} {11}},\ \bibinfo {pages} {974} (\bibinfo {year} {2020})}\BibitemShut {NoStop}%
\bibitem [{\citenamefont {Schweizer}\ \emph {et~al.}(2016)\citenamefont {Schweizer}, \citenamefont {Lohse}, \citenamefont {Citro},\ and\ \citenamefont {Bloch}}]{schweizerSpinPumpingMeasurement2016}%
  \BibitemOpen
  \bibfield  {author} {\bibinfo {author} {\bibfnamefont {C.}~\bibnamefont {Schweizer}}, \bibinfo {author} {\bibfnamefont {M.}~\bibnamefont {Lohse}}, \bibinfo {author} {\bibfnamefont {R.}~\bibnamefont {Citro}},\ and\ \bibinfo {author} {\bibfnamefont {I.}~\bibnamefont {Bloch}},\ }\bibfield  {title} {\bibinfo {title} {Spin {{Pumping}} and {{Measurement}} of {{Spin Currents}} in {{Optical Superlattices}}},\ }\href {https://doi.org/10.1103/PhysRevLett.117.170405} {\bibfield  {journal} {\bibinfo  {journal} {Physical Review Letters}\ }\textbf {\bibinfo {volume} {117}},\ \bibinfo {pages} {170405} (\bibinfo {year} {2016})}\BibitemShut {NoStop}%
\bibitem [{\citenamefont {Cerjan}\ \emph {et~al.}(2020)\citenamefont {Cerjan}, \citenamefont {Wang}, \citenamefont {Huang}, \citenamefont {Chen},\ and\ \citenamefont {Rechtsman}}]{cerjanThoulessPumpingDisordered2020}%
  \BibitemOpen
  \bibfield  {author} {\bibinfo {author} {\bibfnamefont {A.}~\bibnamefont {Cerjan}}, \bibinfo {author} {\bibfnamefont {M.}~\bibnamefont {Wang}}, \bibinfo {author} {\bibfnamefont {S.}~\bibnamefont {Huang}}, \bibinfo {author} {\bibfnamefont {K.~P.}\ \bibnamefont {Chen}},\ and\ \bibinfo {author} {\bibfnamefont {M.~C.}\ \bibnamefont {Rechtsman}},\ }\bibfield  {title} {\bibinfo {title} {Thouless pumping in disordered photonic systems},\ }\href {https://doi.org/10.1038/s41377-020-00408-2} {\bibfield  {journal} {\bibinfo  {journal} {Light: Science \& Applications}\ }\textbf {\bibinfo {volume} {9}},\ \bibinfo {pages} {178} (\bibinfo {year} {2020})}\BibitemShut {NoStop}%
\bibitem [{\citenamefont {Nakajima}\ \emph {et~al.}(2021)\citenamefont {Nakajima}, \citenamefont {Takei}, \citenamefont {Sakuma}, \citenamefont {Kuno}, \citenamefont {Marra},\ and\ \citenamefont {Takahashi}}]{nakajimaCompetitionInterplayTopology2021}%
  \BibitemOpen
  \bibfield  {author} {\bibinfo {author} {\bibfnamefont {S.}~\bibnamefont {Nakajima}}, \bibinfo {author} {\bibfnamefont {N.}~\bibnamefont {Takei}}, \bibinfo {author} {\bibfnamefont {K.}~\bibnamefont {Sakuma}}, \bibinfo {author} {\bibfnamefont {Y.}~\bibnamefont {Kuno}}, \bibinfo {author} {\bibfnamefont {P.}~\bibnamefont {Marra}},\ and\ \bibinfo {author} {\bibfnamefont {Y.}~\bibnamefont {Takahashi}},\ }\bibfield  {title} {\bibinfo {title} {Competition and interplay between topology and quasi-periodic disorder in {{Thouless}} pumping of ultracold atoms},\ }\href {https://doi.org/10.1038/s41567-021-01229-9} {\bibfield  {journal} {\bibinfo  {journal} {Nature Physics}\ }\textbf {\bibinfo {volume} {17}},\ \bibinfo {pages} {844} (\bibinfo {year} {2021})}\BibitemShut {NoStop}%
\bibitem [{\citenamefont {J{\"u}rgensen}\ \emph {et~al.}(2021)\citenamefont {J{\"u}rgensen}, \citenamefont {Mukherjee},\ and\ \citenamefont {Rechtsman}}]{jurgensenQuantizedNonlinearThouless2021}%
  \BibitemOpen
  \bibfield  {author} {\bibinfo {author} {\bibfnamefont {M.}~\bibnamefont {J{\"u}rgensen}}, \bibinfo {author} {\bibfnamefont {S.}~\bibnamefont {Mukherjee}},\ and\ \bibinfo {author} {\bibfnamefont {M.~C.}\ \bibnamefont {Rechtsman}},\ }\bibfield  {title} {\bibinfo {title} {Quantized nonlinear {{Thouless}} pumping},\ }\href {https://doi.org/10.1038/s41586-021-03688-9} {\bibfield  {journal} {\bibinfo  {journal} {Nature}\ }\textbf {\bibinfo {volume} {596}},\ \bibinfo {pages} {63} (\bibinfo {year} {2021})}\BibitemShut {NoStop}%
\bibitem [{\citenamefont {J{\"u}rgensen}\ \emph {et~al.}(2023)\citenamefont {J{\"u}rgensen}, \citenamefont {Mukherjee}, \citenamefont {J{\"o}rg},\ and\ \citenamefont {Rechtsman}}]{jurgensenQuantizedFractionalThouless2023}%
  \BibitemOpen
  \bibfield  {author} {\bibinfo {author} {\bibfnamefont {M.}~\bibnamefont {J{\"u}rgensen}}, \bibinfo {author} {\bibfnamefont {S.}~\bibnamefont {Mukherjee}}, \bibinfo {author} {\bibfnamefont {C.}~\bibnamefont {J{\"o}rg}},\ and\ \bibinfo {author} {\bibfnamefont {M.~C.}\ \bibnamefont {Rechtsman}},\ }\bibfield  {title} {\bibinfo {title} {Quantized fractional {{Thouless}} pumping of solitons},\ }\href {https://doi.org/10.1038/s41567-022-01871-x} {\bibfield  {journal} {\bibinfo  {journal} {Nature Physics}\ }\textbf {\bibinfo {volume} {19}},\ \bibinfo {pages} {420} (\bibinfo {year} {2023})}\BibitemShut {NoStop}%
\bibitem [{\citenamefont {Walter}\ \emph {et~al.}(2023)\citenamefont {Walter}, \citenamefont {Zhu}, \citenamefont {G{\"a}chter}, \citenamefont {Minguzzi}, \citenamefont {Roschinski}, \citenamefont {Sandholzer}, \citenamefont {Viebahn},\ and\ \citenamefont {Esslinger}}]{walterQuantizationItsBreakdown2023}%
  \BibitemOpen
  \bibfield  {author} {\bibinfo {author} {\bibfnamefont {A.-S.}\ \bibnamefont {Walter}}, \bibinfo {author} {\bibfnamefont {Z.}~\bibnamefont {Zhu}}, \bibinfo {author} {\bibfnamefont {M.}~\bibnamefont {G{\"a}chter}}, \bibinfo {author} {\bibfnamefont {J.}~\bibnamefont {Minguzzi}}, \bibinfo {author} {\bibfnamefont {S.}~\bibnamefont {Roschinski}}, \bibinfo {author} {\bibfnamefont {K.}~\bibnamefont {Sandholzer}}, \bibinfo {author} {\bibfnamefont {K.}~\bibnamefont {Viebahn}},\ and\ \bibinfo {author} {\bibfnamefont {T.}~\bibnamefont {Esslinger}},\ }\bibfield  {title} {\bibinfo {title} {Quantization and its breakdown in a {{Hubbard}}--{{Thouless}} pump},\ }\href@noop {} {\bibfield  {journal} {\bibinfo  {journal} {Nature Physics}\ ,\ \bibinfo {pages} {1}} (\bibinfo {year} {2023})}\BibitemShut {NoStop}%
\bibitem [{\citenamefont {Viebahn}\ \emph {et~al.}(2023)\citenamefont {Viebahn}, \citenamefont {Walter}, \citenamefont {Bertok}, \citenamefont {Zhu}, \citenamefont {G{\"a}chter}, \citenamefont {Aligia}, \citenamefont {{Heidrich-Meisner}},\ and\ \citenamefont {Esslinger}}]{viebahnInteractioninducedChargePumping2023}%
  \BibitemOpen
  \bibfield  {author} {\bibinfo {author} {\bibfnamefont {K.}~\bibnamefont {Viebahn}}, \bibinfo {author} {\bibfnamefont {A.-S.}\ \bibnamefont {Walter}}, \bibinfo {author} {\bibfnamefont {E.}~\bibnamefont {Bertok}}, \bibinfo {author} {\bibfnamefont {Z.}~\bibnamefont {Zhu}}, \bibinfo {author} {\bibfnamefont {M.}~\bibnamefont {G{\"a}chter}}, \bibinfo {author} {\bibfnamefont {A.~A.}\ \bibnamefont {Aligia}}, \bibinfo {author} {\bibfnamefont {F.}~\bibnamefont {{Heidrich-Meisner}}},\ and\ \bibinfo {author} {\bibfnamefont {T.}~\bibnamefont {Esslinger}},\ }\bibfield  {title} {\bibinfo {title} {Interaction-induced charge pumping in a topological many-body system},\ }\href@noop {} {\bibfield  {journal} {\bibinfo  {journal} {arXiv preprint arXiv:2308.03756}\ } (\bibinfo {year} {2023})},\ \Eprint {https://arxiv.org/abs/2308.03756} {arxiv:2308.03756} \BibitemShut {NoStop}%
\bibitem [{\citenamefont {Martin}\ \emph {et~al.}(2017)\citenamefont {Martin}, \citenamefont {Refael},\ and\ \citenamefont {Halperin}}]{martinTopologicalFrequencyConversion2017}%
  \BibitemOpen
  \bibfield  {author} {\bibinfo {author} {\bibfnamefont {I.}~\bibnamefont {Martin}}, \bibinfo {author} {\bibfnamefont {G.}~\bibnamefont {Refael}},\ and\ \bibinfo {author} {\bibfnamefont {B.}~\bibnamefont {Halperin}},\ }\bibfield  {title} {\bibinfo {title} {Topological {{Frequency Conversion}} in {{Strongly Driven Quantum Systems}}},\ }\href {https://doi.org/10.1103/PhysRevX.7.041008} {\bibfield  {journal} {\bibinfo  {journal} {Physical Review X}\ }\textbf {\bibinfo {volume} {7}},\ \bibinfo {pages} {041008} (\bibinfo {year} {2017})}\BibitemShut {NoStop}%
\bibitem [{\citenamefont {Fedorova}\ \emph {et~al.}(2020)\citenamefont {Fedorova}, \citenamefont {Qiu}, \citenamefont {Linden},\ and\ \citenamefont {Kroha}}]{fedorovaObservationTopologicalTransport2020}%
  \BibitemOpen
  \bibfield  {author} {\bibinfo {author} {\bibfnamefont {Z.}~\bibnamefont {Fedorova}}, \bibinfo {author} {\bibfnamefont {H.}~\bibnamefont {Qiu}}, \bibinfo {author} {\bibfnamefont {S.}~\bibnamefont {Linden}},\ and\ \bibinfo {author} {\bibfnamefont {J.}~\bibnamefont {Kroha}},\ }\bibfield  {title} {\bibinfo {title} {Observation of topological transport quantization by dissipation in fast {{Thouless}} pumps},\ }\href {https://doi.org/10.1038/s41467-020-17510-z} {\bibfield  {journal} {\bibinfo  {journal} {Nature Communications}\ }\textbf {\bibinfo {volume} {11}},\ \bibinfo {pages} {3758} (\bibinfo {year} {2020})}\BibitemShut {NoStop}%
\bibitem [{\citenamefont {Dreon}\ \emph {et~al.}(2022)\citenamefont {Dreon}, \citenamefont {Baumg{\"a}rtner}, \citenamefont {Li}, \citenamefont {Hertlein}, \citenamefont {Esslinger},\ and\ \citenamefont {Donner}}]{dreonSelfoscillatingPumpTopological2022}%
  \BibitemOpen
  \bibfield  {author} {\bibinfo {author} {\bibfnamefont {D.}~\bibnamefont {Dreon}}, \bibinfo {author} {\bibfnamefont {A.}~\bibnamefont {Baumg{\"a}rtner}}, \bibinfo {author} {\bibfnamefont {X.}~\bibnamefont {Li}}, \bibinfo {author} {\bibfnamefont {S.}~\bibnamefont {Hertlein}}, \bibinfo {author} {\bibfnamefont {T.}~\bibnamefont {Esslinger}},\ and\ \bibinfo {author} {\bibfnamefont {T.}~\bibnamefont {Donner}},\ }\bibfield  {title} {\bibinfo {title} {Self-oscillating pump in a topological dissipative atom--cavity system},\ }\href {https://doi.org/10.1038/s41586-022-04970-0} {\bibfield  {journal} {\bibinfo  {journal} {Nature}\ }\textbf {\bibinfo {volume} {608}},\ \bibinfo {pages} {494} (\bibinfo {year} {2022})}\BibitemShut {NoStop}%
\bibitem [{\citenamefont {Sridhar}\ \emph {et~al.}(2024)\citenamefont {Sridhar}, \citenamefont {Ghosh}, \citenamefont {Srinivasan}, \citenamefont {Miller},\ and\ \citenamefont {Dutt}}]{sridharQuantizedTopologicalPumping2024}%
  \BibitemOpen
  \bibfield  {author} {\bibinfo {author} {\bibfnamefont {S.~K.}\ \bibnamefont {Sridhar}}, \bibinfo {author} {\bibfnamefont {S.}~\bibnamefont {Ghosh}}, \bibinfo {author} {\bibfnamefont {D.}~\bibnamefont {Srinivasan}}, \bibinfo {author} {\bibfnamefont {A.~R.}\ \bibnamefont {Miller}},\ and\ \bibinfo {author} {\bibfnamefont {A.}~\bibnamefont {Dutt}},\ }\bibfield  {title} {\bibinfo {title} {Quantized topological pumping in {{Floquet}} synthetic dimensions with a driven dissipative photonic molecule},\ }\bibfield  {journal} {\bibinfo  {journal} {Nature Physics}\ }\href {https://doi.org/10.1038/s41567-024-02413-3} {10.1038/s41567-024-02413-3} (\bibinfo {year} {2024})\BibitemShut {NoStop}%
\bibitem [{\citenamefont {Hasan}\ and\ \citenamefont {Kane}(2010)}]{hasanColloquiumTopologicalInsulators2010}%
  \BibitemOpen
  \bibfield  {author} {\bibinfo {author} {\bibfnamefont {M.~Z.}\ \bibnamefont {Hasan}}\ and\ \bibinfo {author} {\bibfnamefont {C.~L.}\ \bibnamefont {Kane}},\ }\bibfield  {title} {\bibinfo {title} {Colloquium: Topological insulators},\ }\href@noop {} {\bibfield  {journal} {\bibinfo  {journal} {Reviews of Modern Physics}\ }\textbf {\bibinfo {volume} {82}},\ \bibinfo {pages} {3045} (\bibinfo {year} {2010})}\BibitemShut {NoStop}%
\bibitem [{\citenamefont {Cooper}\ \emph {et~al.}(2019)\citenamefont {Cooper}, \citenamefont {Dalibard},\ and\ \citenamefont {Spielman}}]{cooperTopologicalBandsUltracold2019}%
  \BibitemOpen
  \bibfield  {author} {\bibinfo {author} {\bibfnamefont {N.~R.}\ \bibnamefont {Cooper}}, \bibinfo {author} {\bibfnamefont {J.}~\bibnamefont {Dalibard}},\ and\ \bibinfo {author} {\bibfnamefont {I.~B.}\ \bibnamefont {Spielman}},\ }\bibfield  {title} {\bibinfo {title} {Topological bands for ultracold atoms},\ }\href {https://doi.org/10.1103/RevModPhys.91.015005} {\bibfield  {journal} {\bibinfo  {journal} {Reviews of Modern Physics}\ }\textbf {\bibinfo {volume} {91}},\ \bibinfo {pages} {015005} (\bibinfo {year} {2019})}\BibitemShut {NoStop}%
\bibitem [{\citenamefont {Ozawa}\ \emph {et~al.}(2019)\citenamefont {Ozawa}, \citenamefont {Price}, \citenamefont {Amo}, \citenamefont {Goldman}, \citenamefont {Hafezi}, \citenamefont {Lu}, \citenamefont {Rechtsman}, \citenamefont {Schuster}, \citenamefont {Simon}, \citenamefont {Zilberberg},\ and\ \citenamefont {Carusotto}}]{ozawaTopologicalPhotonics2019}%
  \BibitemOpen
  \bibfield  {author} {\bibinfo {author} {\bibfnamefont {T.}~\bibnamefont {Ozawa}}, \bibinfo {author} {\bibfnamefont {H.~M.}\ \bibnamefont {Price}}, \bibinfo {author} {\bibfnamefont {A.}~\bibnamefont {Amo}}, \bibinfo {author} {\bibfnamefont {N.}~\bibnamefont {Goldman}}, \bibinfo {author} {\bibfnamefont {M.}~\bibnamefont {Hafezi}}, \bibinfo {author} {\bibfnamefont {L.}~\bibnamefont {Lu}}, \bibinfo {author} {\bibfnamefont {M.~C.}\ \bibnamefont {Rechtsman}}, \bibinfo {author} {\bibfnamefont {D.}~\bibnamefont {Schuster}}, \bibinfo {author} {\bibfnamefont {J.}~\bibnamefont {Simon}}, \bibinfo {author} {\bibfnamefont {O.}~\bibnamefont {Zilberberg}},\ and\ \bibinfo {author} {\bibfnamefont {I.}~\bibnamefont {Carusotto}},\ }\bibfield  {title} {\bibinfo {title} {Topological photonics},\ }\href {https://doi.org/10.1103/RevModPhys.91.015006} {\bibfield  {journal} {\bibinfo  {journal} {Reviews of Modern Physics}\ }\textbf {\bibinfo {volume} {91}},\ \bibinfo {pages} {015006} (\bibinfo {year} {2019})}\BibitemShut {NoStop}%
\bibitem [{\citenamefont {Nash}\ \emph {et~al.}(2015)\citenamefont {Nash}, \citenamefont {Kleckner}, \citenamefont {Read}, \citenamefont {Vitelli}, \citenamefont {Turner},\ and\ \citenamefont {Irvine}}]{nashTopologicalMechanicsGyroscopic2015}%
  \BibitemOpen
  \bibfield  {author} {\bibinfo {author} {\bibfnamefont {L.~M.}\ \bibnamefont {Nash}}, \bibinfo {author} {\bibfnamefont {D.}~\bibnamefont {Kleckner}}, \bibinfo {author} {\bibfnamefont {A.}~\bibnamefont {Read}}, \bibinfo {author} {\bibfnamefont {V.}~\bibnamefont {Vitelli}}, \bibinfo {author} {\bibfnamefont {A.~M.}\ \bibnamefont {Turner}},\ and\ \bibinfo {author} {\bibfnamefont {W.~T.~M.}\ \bibnamefont {Irvine}},\ }\bibfield  {title} {\bibinfo {title} {Topological mechanics of gyroscopic metamaterials},\ }\href {https://doi.org/10.1073/pnas.1507413112} {\bibfield  {journal} {\bibinfo  {journal} {Proceedings of the National Academy of Sciences}\ }\textbf {\bibinfo {volume} {112}},\ \bibinfo {pages} {14495} (\bibinfo {year} {2015})}\BibitemShut {NoStop}%
\bibitem [{\citenamefont {Huber}(2016)}]{huberTopologicalMechanics2016}%
  \BibitemOpen
  \bibfield  {author} {\bibinfo {author} {\bibfnamefont {S.~D.}\ \bibnamefont {Huber}},\ }\bibfield  {title} {\bibinfo {title} {Topological mechanics},\ }\href {https://doi.org/10.1038/nphys3801} {\bibfield  {journal} {\bibinfo  {journal} {Nature Physics}\ }\textbf {\bibinfo {volume} {12}},\ \bibinfo {pages} {621} (\bibinfo {year} {2016})}\BibitemShut {NoStop}%
\bibitem [{\citenamefont {Braun}\ and\ \citenamefont {Kivshar}(1998)}]{braunNonlinearDynamicsFrenkel1998}%
  \BibitemOpen
  \bibfield  {author} {\bibinfo {author} {\bibfnamefont {O.~M.}\ \bibnamefont {Braun}}\ and\ \bibinfo {author} {\bibfnamefont {Y.~S.}\ \bibnamefont {Kivshar}},\ }\bibfield  {title} {\bibinfo {title} {Nonlinear dynamics of the {{Frenkel}}--{{Kontorova}} model},\ }\href {https://doi.org/10.1016/S0370-1573(98)00029-5} {\bibfield  {journal} {\bibinfo  {journal} {Physics Reports}\ }\textbf {\bibinfo {volume} {306}},\ \bibinfo {pages} {1} (\bibinfo {year} {1998})}\BibitemShut {NoStop}%
\bibitem [{\citenamefont {Kivshar}\ and\ \citenamefont {Campbell}(1993)}]{kivsharPeierlsNabarroPotentialBarrier1993}%
  \BibitemOpen
  \bibfield  {author} {\bibinfo {author} {\bibfnamefont {Y.~S.}\ \bibnamefont {Kivshar}}\ and\ \bibinfo {author} {\bibfnamefont {D.~K.}\ \bibnamefont {Campbell}},\ }\bibfield  {title} {\bibinfo {title} {Peierls-{{Nabarro}} potential barrier for highly localized nonlinear modes},\ }\href {https://doi.org/10.1103/PhysRevE.48.3077} {\bibfield  {journal} {\bibinfo  {journal} {Physical Review E}\ }\textbf {\bibinfo {volume} {48}},\ \bibinfo {pages} {3077} (\bibinfo {year} {1993})}\BibitemShut {NoStop}%
\bibitem [{\citenamefont {Harper}(1955)}]{harperSingleBandMotion1955}%
  \BibitemOpen
  \bibfield  {author} {\bibinfo {author} {\bibfnamefont {P.~G.}\ \bibnamefont {Harper}},\ }\bibfield  {title} {\bibinfo {title} {Single band motion of conduction electrons in a uniform magnetic field},\ }\href@noop {} {\bibfield  {journal} {\bibinfo  {journal} {Proceedings of the Physical Society. Section A}\ }\textbf {\bibinfo {volume} {68}},\ \bibinfo {pages} {874} (\bibinfo {year} {1955})}\BibitemShut {NoStop}%
\bibitem [{\citenamefont {Aubry}\ and\ \citenamefont {Andr{\'e}}(1980)}]{aubryAnalyticityBreakingAnderson1980}%
  \BibitemOpen
  \bibfield  {author} {\bibinfo {author} {\bibfnamefont {S.}~\bibnamefont {Aubry}}\ and\ \bibinfo {author} {\bibfnamefont {G.}~\bibnamefont {Andr{\'e}}},\ }\bibfield  {title} {\bibinfo {title} {Analyticity breaking and {{Anderson}} localization in incommensurate lattices},\ }\href@noop {} {\bibfield  {journal} {\bibinfo  {journal} {Ann. Israel Phys. Soc}\ }\textbf {\bibinfo {volume} {3}},\ \bibinfo {pages} {18} (\bibinfo {year} {1980})}\BibitemShut {NoStop}%
\bibitem [{\citenamefont {Süsstrunk}\ and\ \citenamefont {Huber}(2016)}]{suesstrunkClassification2016}%
  \BibitemOpen
  \bibfield  {author} {\bibinfo {author} {\bibfnamefont {R.}~\bibnamefont {Süsstrunk}}\ and\ \bibinfo {author} {\bibfnamefont {S.~D.}\ \bibnamefont {Huber}},\ }\bibfield  {title} {\bibinfo {title} {Classification of topological phonons in linear mechanical metamaterials},\ }\href {https://doi.org/10.1073/pnas.1605462113} {\bibfield  {journal} {\bibinfo  {journal} {Proceedings of the National Academy of Sciences}\ }\textbf {\bibinfo {volume} {113}},\ \bibinfo {pages} {E4767} (\bibinfo {year} {2016})},\ \Eprint {https://arxiv.org/abs/https://www.pnas.org/doi/pdf/10.1073/pnas.1605462113} {https://www.pnas.org/doi/pdf/10.1073/pnas.1605462113} \BibitemShut {NoStop}%
\bibitem [{\citenamefont {Rice}\ and\ \citenamefont {Mele}(1982)}]{riceElementary1982}%
  \BibitemOpen
  \bibfield  {author} {\bibinfo {author} {\bibfnamefont {M.~J.}\ \bibnamefont {Rice}}\ and\ \bibinfo {author} {\bibfnamefont {E.~J.}\ \bibnamefont {Mele}},\ }\bibfield  {title} {\bibinfo {title} {Elementary excitations of a linearly conjugated diatomic polymer},\ }\href {https://doi.org/10.1103/PhysRevLett.49.1455} {\bibfield  {journal} {\bibinfo  {journal} {Phys. Rev. Lett.}\ }\textbf {\bibinfo {volume} {49}},\ \bibinfo {pages} {1455} (\bibinfo {year} {1982})}\BibitemShut {NoStop}%
\end{thebibliography}%

\appendix 
\begin{widetext}

\section{Fabrication \& Materials \& Setup}
The pendulums, the airflow directors attached to the fans, and the torsion springs are 3D printed using a Prusa Mini+ printer and polylactic acid (PLA) filament. To reduce friction, the pendulums are mounted onto a chrome plated carbon steel rod (10mm diameter) using bearings (uxcell 6700-2RS). We use 120mm-large centrifugal fans (GDSTIME) with a nominal airflow of 35cfm, powered by a 12V power supply (MEAN WELL SE-600-12). To create the pumping sequences, every third fan is wired together and controlled using a 25kHz pulse-width modulation signal coming from a single ATmega328P. We evaluate the Thouless pumping by taking videos using a camera (Google Pixel 4a) placed above the system facing downwards.


\section{Characterization}

We characterize the properties of the pendulums, springs and fans by measuring the response of individual elements instead of the whole system. From these measurements, we extract the relevant experimental parameters, like the the spring constant, the friction, the strength of the fans and the moment of inertia. We measure the mass of a pendulum (including the bearing and the screw) to be $M=7.2$\,g. We estimate the center-of-mass distance from the rotation axis to be $L=2.6$\,cm. Each nut has a weight of $m_{\text{nut}}=0.9$\,g and is attached about $l=5.9$\,cm away from the rotation axis.

We evaluate the strength of friction and the fan's influence by monitoring the damped oscillation of a single pendulum. This is shown in Fig. \ref{Fig:Characterization}a for a pendulum with additional weight $m$ of two nuts attached at distance $l$, while the fan below is turned off (blue) or on (red). Corresponding numerical simulations using the equation of motion for a single pendulum,
\begin{equation}
    \ddot{\phi} = -D \sin\phi - \alpha \dot{\phi},
\end{equation}
where $D=\frac{MgL+mgl}{I}$, are shown in black and used to extract the strength of the friction, the moment of inertia, as well as the strength of the fans. In the case of switched-off fans, the simulations use a moment of inertia $I_{\text{off}}=1.6 M L^2 + m l^2$ and $\alpha=2.0$/s, where the 1.6 is a correction factor to account for the pendulum's shape. In the case of switched-on fans, the simulations uses $D_{\text{on}} = 0.45 D_{\text{off}}$ and a slightly higher friction value of $\alpha=2.4$/s. Differently stated, each fan can counteract 45\% of the gravitational force (and hence the potential energy) of a single pendulum with two attached nuts, or about 85\% of a pendulum without attached nuts. Similarly, we estimate the friction of pendulums without additional nuts to be $\alpha=4.2$/s.

\begin{figure}[htbp]
    \includegraphics[]{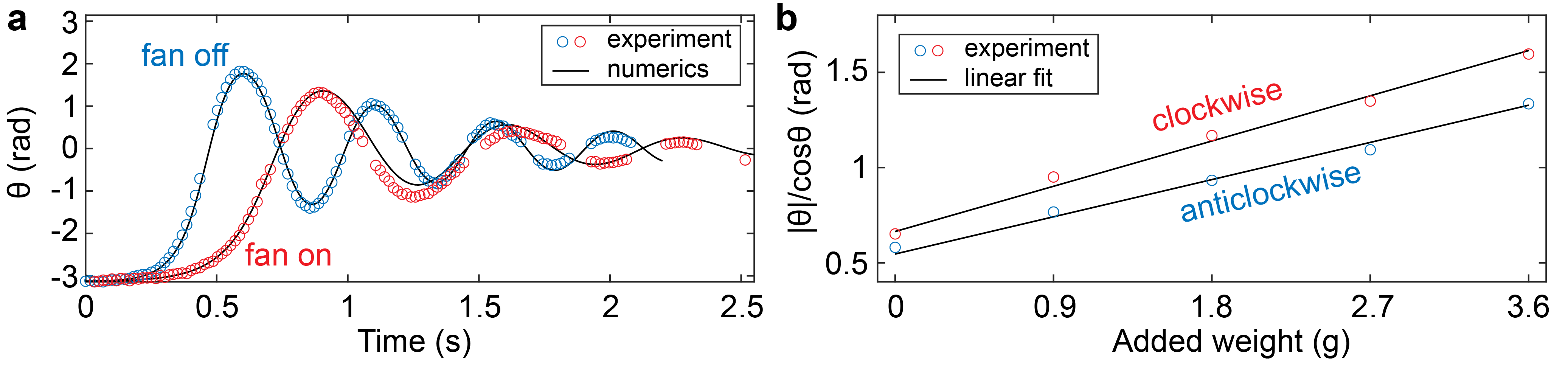}
    \caption{\textbf{Characterization of pendulum, spring and fan components.} \textbf{a}. Measured damped oscillations of a single pendulum when the fan below is switched off (blue) and on (red). Corresponding numerical simulations are shown in black. \textbf{b} Measured opening angle $\theta$ of two pendulums connected by a single torsion spring as a function of attached additional weight. Black lines are linear fits with a slope inversely proportional to the torsion spring constant.}
    \label{Fig:Characterization}
\end{figure}

We determine the spring constant of our torsion springs using a minimal system consisting of one torsion spring connected to two pendulums. The first pendulum is held approximately horizontally. The second pendulum is free to rotate. We measure the opening angle $\theta$ as a function of increasing weights $m$ attached to the pendulum's end (at position $l$) for clockwise and anticlockwise rotation. 

Using the definition of the torsion spring constant
\begin{equation}
    k = \frac{|\vec{\tau}|}{\theta} = \frac{\vec{F}_g \cross \vec{r}}{\theta} = \frac{(MgL+mgl) \cos\theta}{\theta}
\end{equation}
where $\vec{\tau}$ is the applied torque, $\vec{F}_g$ the gravitational force, $M$ the mass of the bare pendulum, $L$ the center-of-mass distance of the bare pendulum from the rotation axis, and $g$ the gravitational acceleration.

Solving for
\begin{equation}
    \frac{\theta}{\cos{\theta}} = \frac{(MgL+mgl)}{k}
\end{equation}
with a subsequent linear fit, as shown in Fig. 6b, results in $k=2.7\frac{\text{mNm}}{\text{rad}}$ for clockwise and $k=2.2\frac{\text{mNm}}{\text{rad}}$ for anti-clockwise rotation.


\section{Data acquisition and handling}
In order to evaluate the position of the pendulums in the system, we take videos from the top viewing downwards onto the bar/pendulums. We track their position using the attached screws, that can be color-masked in the videos. This method allows us to take data around the center of the soliton, but not for its tails, as pendulums that point downwards are hidden behind the bar and the springs. While evaluating the position of the pendulums, we calculate their distance from the bar and convert this into angles. As each distance corresponds to two possible angles, we furthermore rely on our knowledge that the pendulums form a kink soliton to determine the angle.

\section{Lagrangian, Hamiltonian and equations of motion}

We describe our system with the following Lagrangian, that consists of the kinetic (rotational) energy of the pendulums, the potential energy of the pendulums and the energy stored in the torsion springs:
\begin{equation}
    \mathcal{L} (t) = T - V = \sum_n \frac{I_n}{2} \dot{\phi}_n^2 - \sum_n \left( M_n g- F_n(t) \right) L (1-\cos\phi_n) \nonumber \\
    - \sum_n \frac{k}{2} \left( \phi_n-\phi_{n+1} \right) ^2.
\end{equation}
Here, $I_n$ denotes the moment of inertia of the pendulum on site $n$, $M_n$ is the mass of the pendulum on site $n$, $g$ denotes the gravity of earth, $L$ is the center-of-mass distance of the pendulum from its rotation axis, $k$ denotes the torsion spring constant, and we included the effect of the fans as a force $F_n$ counteracting the gravitational force. Using Lagrange's equations, we derive the following equations of motion:
\begin{equation}
    \ddot{\phi}_n = -\frac{(M_n g -F_n) L}{I_n} \sin(\phi_n) + \frac{k}{I_n} \left( \phi_{n+1}+\phi_{n-1}-2 \phi_n \right) + \alpha \dot{\phi}_n,
\end{equation}
where we phenomenologically included a friction term with strength $\alpha$. Using the following definitions
\begin{align}
    D_n(\Omega t) &=  \frac{(M_n g -F_n(\Omega t)) L}{I_n} \\
    K_n &= k/I_n
\end{align}
we arrive at
\begin{equation}
    \ddot{\phi}_n = -D_n \sin(\phi_n) + K_n \left( \phi_{n+1}+\phi_{n-1}-2 \phi_n \right) + \alpha \dot{\phi}_n,
\end{equation}
which is Eq. (1) of the main part.

Similarly, the (classical) Hamiltonian of our system is given via
\begin{equation}
    H(t) = T + V = \sum_n \frac{I_n}{2} \dot{\phi}_n^2 + \sum_n \left( M_n g- F_n \right) L (1-\cos\phi_n)  + \sum_n \frac{k}{2} \left( \phi_n-\phi_{n+1} \right) ^2
\end{equation}

\section{Effect of displaced fans}
In the presence of displaced pendulums, the pendulums on one side of the bar ($0 \leq \phi_n <\pi$) experience a different airflow and hence a different upwards force compared to pendulums on the other side ($\pi \leq \phi_n <2\pi$). To describe these force, we make the following replacement:
\begin{equation}
    F_n \rightarrow \tilde{F}_n = \begin{cases} (1+\beta) F_n  & \mbox{if } 0 \leq \phi_n <\pi \\ (1-\beta) F_n & \mbox{if } \pi \leq \phi_n <2\pi \end{cases}
\end{equation}
where $\beta$ dictates the the difference in the airflow (and hence upwards force) for the two sides of the bar. Furthermore, we have restricted ourselves to $0\leq \phi_n <2\pi$, but the results can be generalized to multiple rotations. 

The corresponding potential energy term in the Hamiltonian that creates this force, can be found via integration. Under these circumstances the Hamiltonian is given by:
\begin{equation}
    H \rightarrow \tilde{H} = H(F_n \rightarrow \tilde{F_n}) + C_n,
\end{equation}
where, $C_n$ are constants that have to be adjusted such that the Hamiltonian is continuous. In other words, in the new Hamiltonian the force $F_n$ is replaced with $\tilde{F}_n$. 

In the main text, we used the variational method and calculated the effective potential for the kink soliton, and numerically adjust the $C_n$, such that $H$ is continuous, leading to a potential gradient as the effective potential landscape.

The effect of the displaced fans can also be understood intuitively for a kink soliton. Suppose the pendulums displacements left of the kink's center are $0\leq\phi_n<\pi$ and hence experience a larger force than the pendulums on the right side, with $\pi<\phi_n<2\pi$. Hence, the pendulums on the left push upwards, while the solitons on the right push downwards, such that the kink's center moves.

We additionally confirm that the fan displacement acts like a potential gradient (equivalent to an electric field for electrons) by simulating the time evolution of a kink soliton at zero friction ($\alpha = 0$) and a spring constant of $k=10$mNm/rad (to decrease friction from the Peierls-Nabarro potential barrier). The results shown in Fig. 7 show that the position of the soliton is well described using a quadratic fit (as known from electrons in electric fields).

\begin{figure}[htbp]
    \includegraphics[]{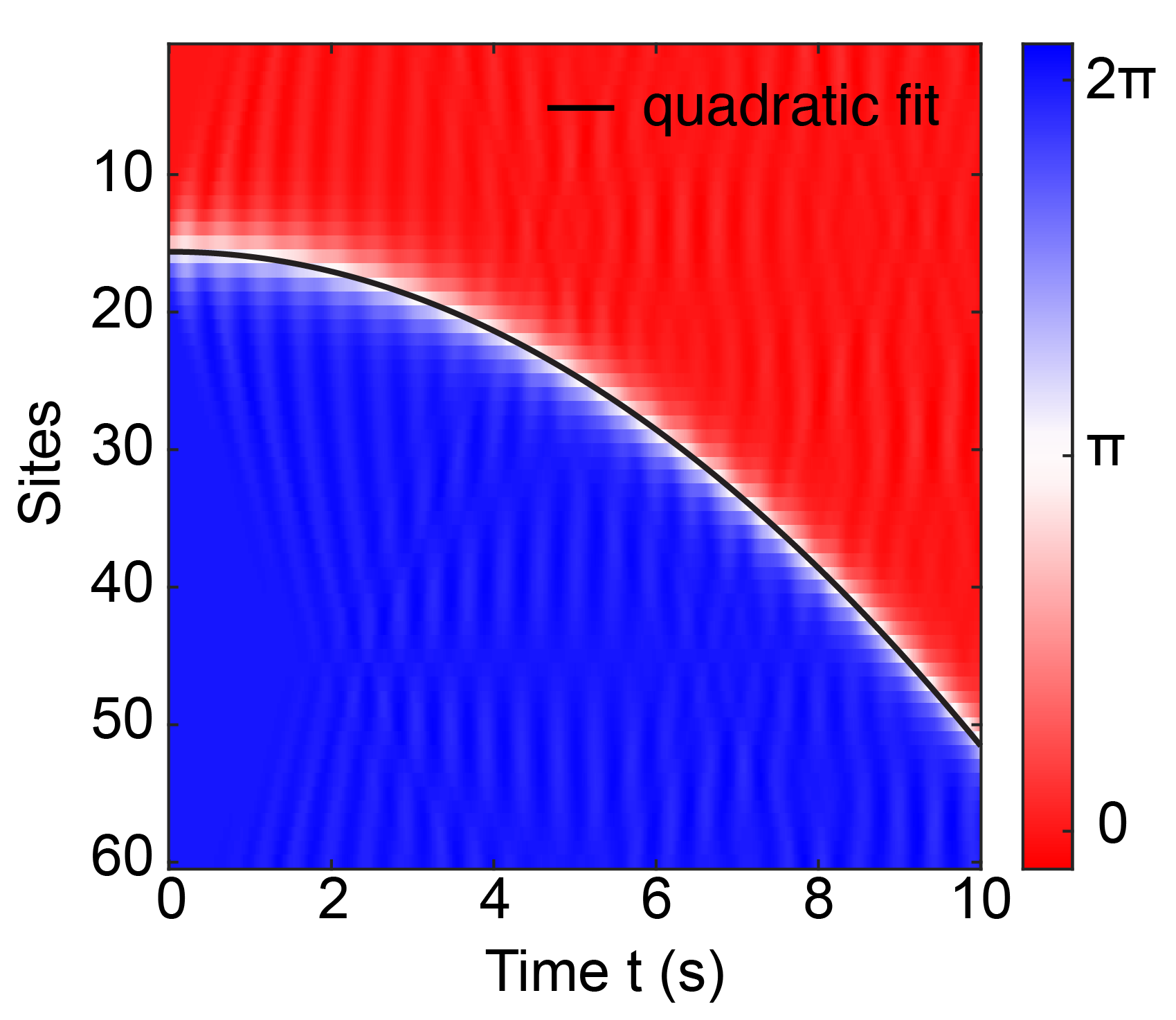}
    \caption{\textbf{Effect of displaced fans.} Time propagation of a kink soliton in a simple lattice (all fans on with equal intensity; $\beta = 0.1$) and no friction ($\alpha=0$). The black line is a quadratic fit to the center of the kink soliton confirming that displaced fans create a potential gradient and the soliton is displaced akin to a charged particle acting in response to an electric field.}
    \label{Fig:E-field}
\end{figure}

\section{Calculation of the topological invariant in the linear, non-dissipative system}

Topological systems are classified by topological invariants which describe the quantized observable. For mechanical systems, this is not different to electronic systems \cite{huberTopologicalMechanics2016}, as long as they are linear and conservative. As our system is nonlinear and dissipative, there is no general procedure to calculate topological invariants.

Instead, we show that the underlying linear, non-dissipative system is topologically non-trivial and described by the Chern number. In order to do so, we follow \cite{suesstrunkClassification2016} and transform the equations of motion into a Schrödinger-like equation, so that the usual toolbox to calculate topological invariants applies. In particular, we calculate the Chern number as the winding of the position of the instantaneous Wannier states around the unit cell during one pump cycle.

\begin{figure}[htbp]
    \includegraphics[]{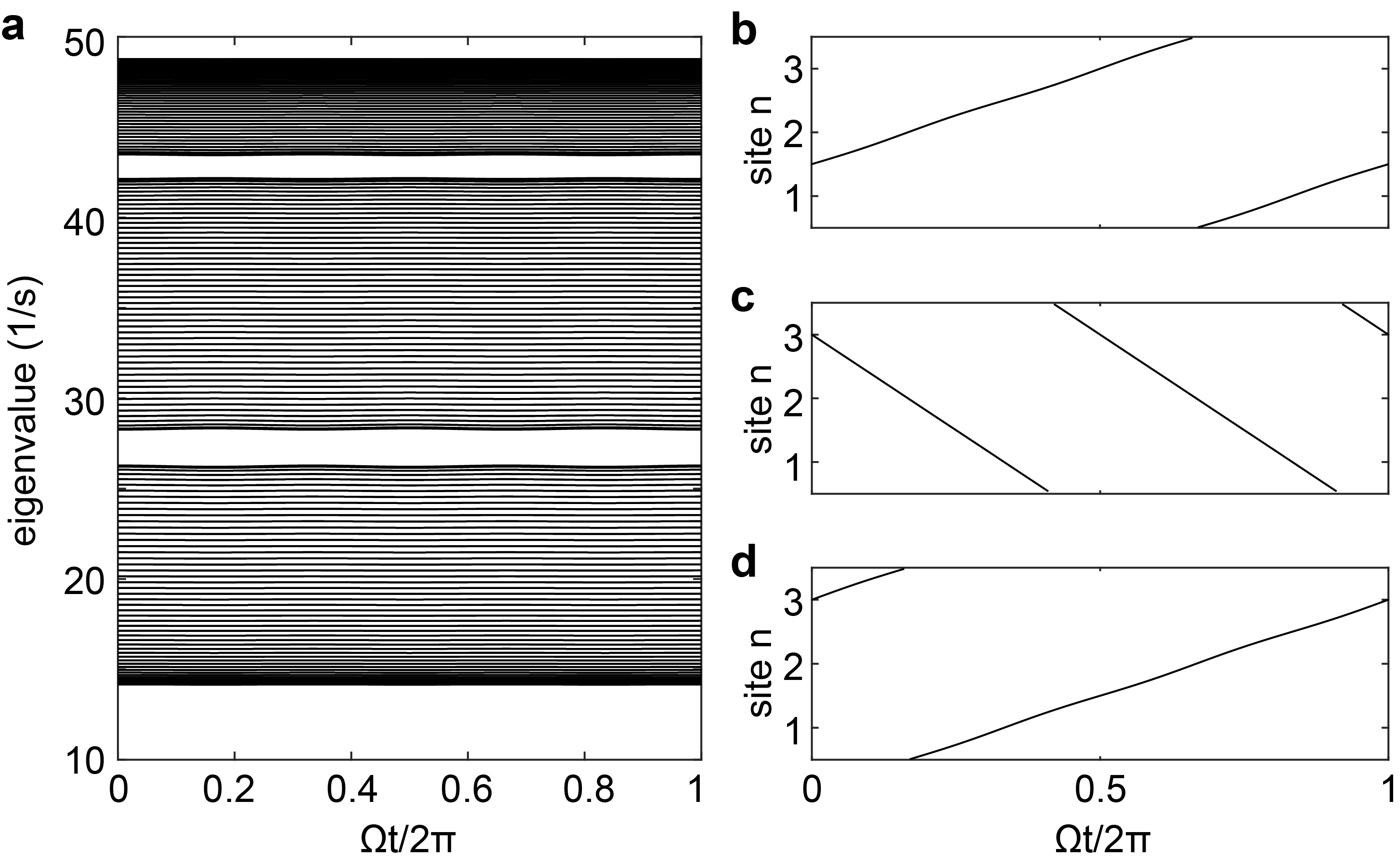}
    \caption{\textbf{Wannier center calculation.} \textbf{a} Calculated band structure as a function of the pump period. \textbf{b-d} Position of the center of the single-band Wannier states of the upper, middle and lower band, respectively. The number of windings around the unit cell defines the Chern number of the respective band.}
    \label{Fig:Wannier}
\end{figure}

Using $\dot{\phi} \equiv y$, $\alpha=0$ and $\sin{\phi_n} \approx \phi_n$ (small oscillations), we can write Eq. (D4) as
\begin{equation}
i \frac{\partial}{\partial t} 
    \begin{pmatrix} \phi \\ y \end{pmatrix} = 
    \begin{pmatrix} 0 & 1 \\ -\mathcal{D}(t) & 0 \end{pmatrix} 
    \begin{pmatrix} \phi \\ y \end{pmatrix}
\end{equation}

where $\phi$ represents the vector of $\phi_n$ and $\mathcal{D}$ represents the dynamical matrix. In order to bring this equation onto the same footing as the Schrödinger equation, we make the following transformation \cite{suesstrunkClassification2016} for the instantaneous system
\begin{equation}
    U = \begin{pmatrix}
    \sqrt{\mathcal{D}} & 0 \\
    0 & i
    \end{pmatrix} 
\end{equation} 
where $\sqrt{\mathcal{D}}$ is defined through its spectral decomposition and the positive branch of the square root of the eigenvalues is chosen. This leads to
\begin{equation}
i \frac{\partial}{\partial t} 
    \begin{pmatrix}
    \sqrt{\mathcal{D}} \phi \\ i y \end{pmatrix} 
    = 
    \begin{pmatrix}
    0 & \sqrt{\mathcal{D}} \\
    \sqrt{\mathcal{D}} & 0
    \end{pmatrix} 
    \begin{pmatrix} \sqrt{\mathcal{D}} \phi \\ i y \end{pmatrix}
\end{equation}
This equation has particle-hole symmetry, hence we only need to block-diagonalize it. The corresponding band structure and the position of the center of the instantaneous Wannier states (calculated using the projected position operator), for each band are shown in Fig.~\ref{Fig:Wannier}. The winding around the unit cell per cycle determines the Chern numbers, that can be read off as +1, -2 and +1 for the upper, middle and center band.

\section{A generalization using the Rice-Mele model}
The observed nonlinear topological pumping is not unique to the particular model chosen in the main part, a three site AAH-model with a discontinuous driving scheme, but is a general phenomena. Here, we show that analogue behavior also occurs in the continuously-modulated Rice-Mele model \cite{riceElementary1982}, perhaps the most prototypical Thouless pumping model, with only two sites per unit cell. In this model, not only the diagonal entries (corresponding to on-site energies in electronic systems) have to be modulated, but also the off-diagonal entries (corresponding to nearest-neighbor hoppings in electronic systems). 

\begin{figure}[htbp]
    \includegraphics[]{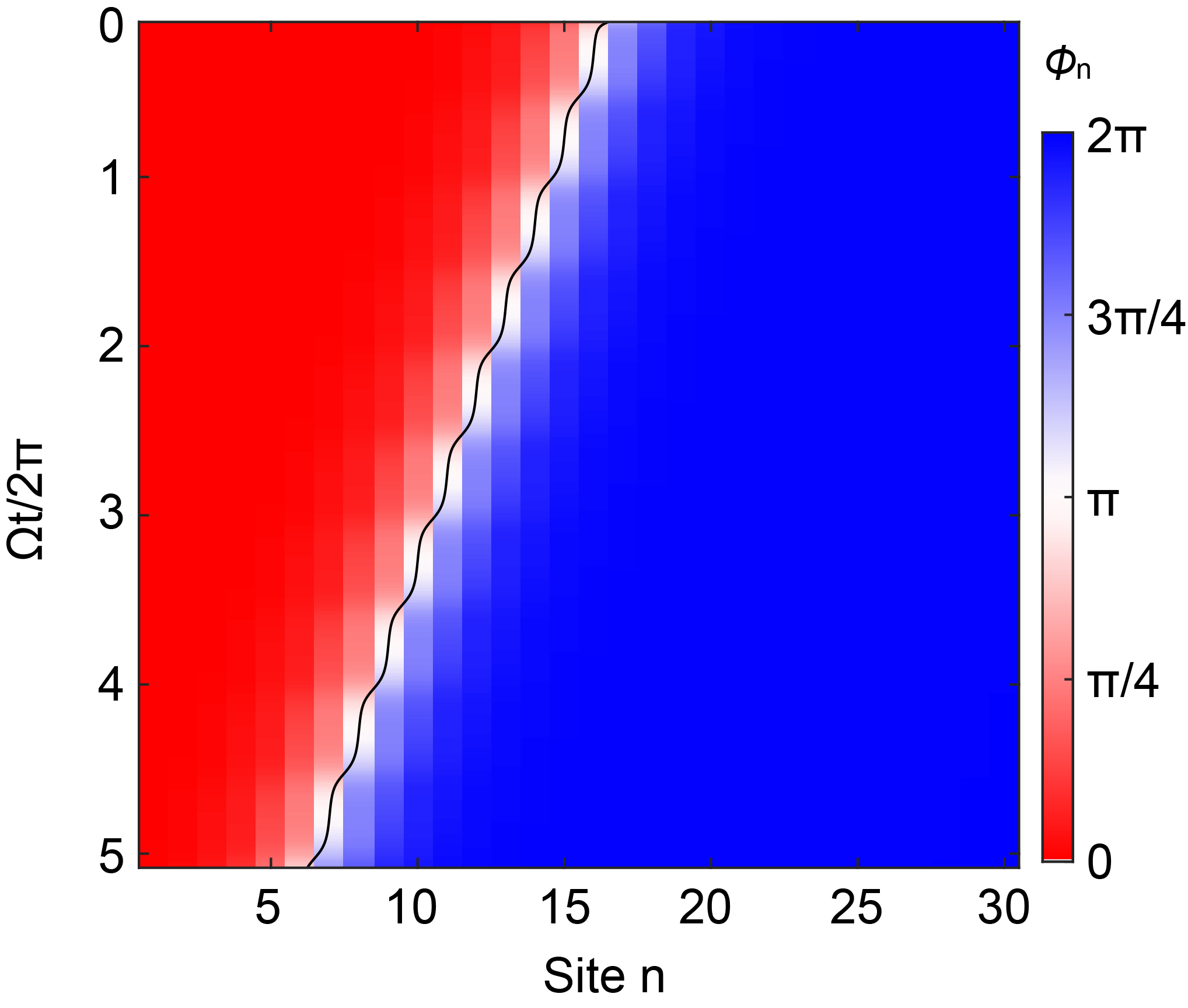}
    \caption{\textbf{Quantized kink soliton pumping in a Rice-Mele model.} The displacement per period is one unit cell (two sites). Parameters: $\alpha=100$/s, $T=100$s.}
    \label{Fig:RiceMele}
\end{figure}

In our model, this means that the spring constants $k_n$ have to be chosen time-dependent. As this is not compatible with our experimental scheme, we only show numerical calculations, using:
\begin{align}
k_n &= k \mp 0.2k \cos{\Omega t}\\
F_n &= \pm 0.57 M_n g \sin{\Omega t}\\
\end{align}
where the upper sign applies to $n$ odd, and the lower to $n$ even.

The corresponding time-dynamics of the kink soliton is shown in Fig.~\ref{Fig:RiceMele}, clearly showing quantized transport of one unit cell per cycle. This confirms that dissipative Thouless pumping of solitons is completely general and works for different pumping models as well as continuous and discontinuous driving protocols.

\section{Additional phase diagram}
Finally, we also show an additional phase diagram to demonstrate the robustness of the quantization of dissipative soliton transport. For a fixed friction ($\alpha=55$/s), we calculate the soliton transport as a function of the strength of the potential gradient, characterized by $\beta$ (see Eq. (E1)). The results are shown in Fig.~\ref{Fig:PhaseDiagramBeta}. As each plateau occurs over a range of values in $T$ and $\beta$, this clearly demonstrates that the observed plateaux of quantization are not a fine-tuned phenomena. Hence, quantization is independent of the exact system parameters and is not perturbed by small perturbations (disorder), the hallmark feature of topological protection.

\begin{figure}[htbp]
    \includegraphics[]{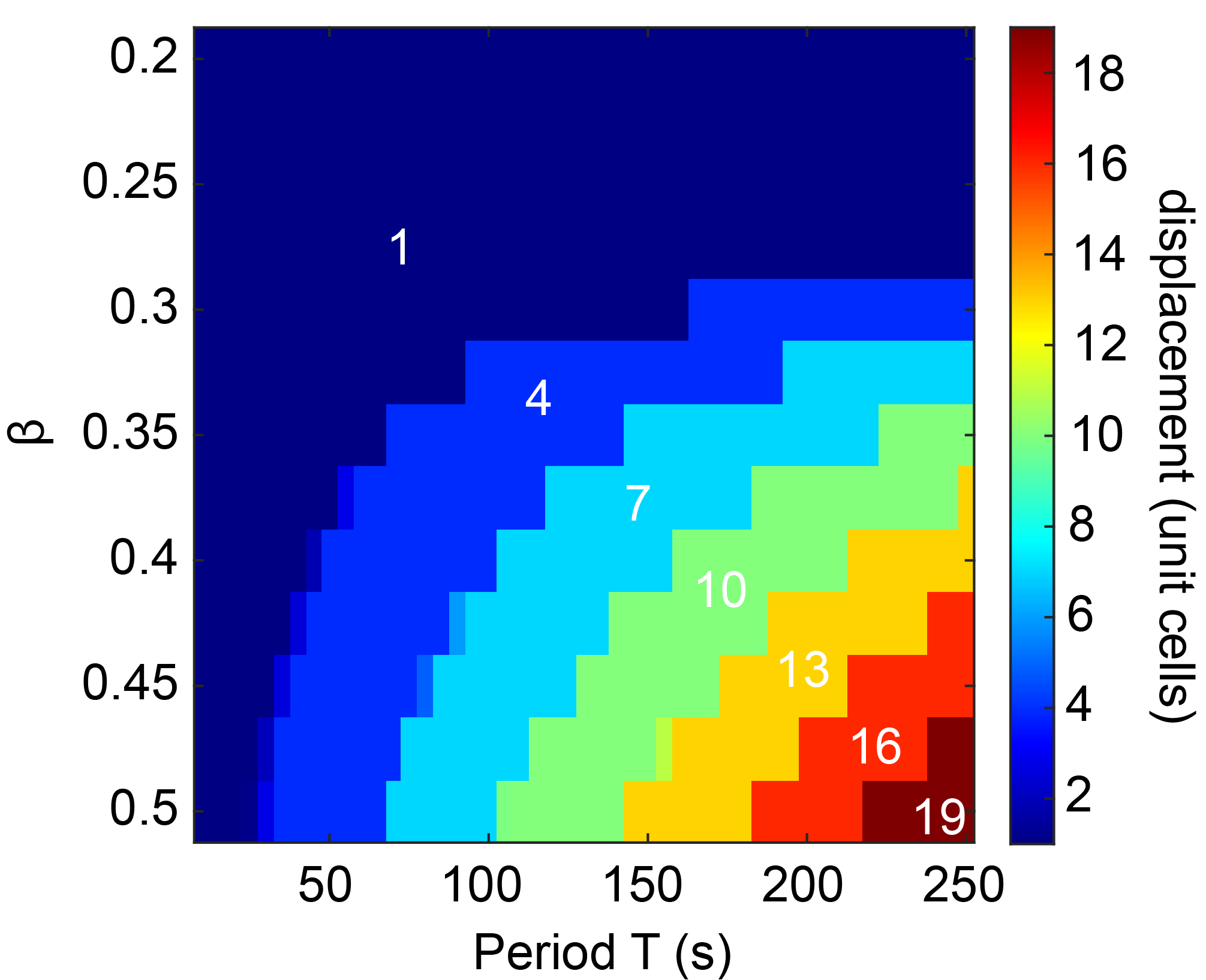}
    \caption{\textbf{Phase diagram as a function of the strength of the potential gradient.} Quantized regimes occur at different pumping speeds (quantified by the period $T$), as a function of the strength of the potential gradient $\beta$. White numbers describe the displacement of the soliton per period in units of the unit cell length.}
    \label{Fig:PhaseDiagramBeta}
\end{figure}

\section{Supplementary Video 1}
Exemplary video that shows the movement of the kink soliton for forward and reverse pumping direction, from which the data of Fig. 2b has been extracted.

\section{Supplementary Animation 1}
This animation shows the position of a forward-pumped kink soliton in its effective potential landscape during one period for $\alpha=\{16,36,100\}$/s, corresponding to a quantized displacement of $\{1,4,7\}$ unit cells. The vertical grey line denotes site 16.


\end{widetext}
\end{document}